\documentclass[12pt]{article}
\usepackage{amsmath,amsthm,amsfonts,amssymb,mathrsfs}
\usepackage{graphicx}
\usepackage[a4paper]{geometry}
\usepackage{array}
\usepackage{enumerate}
\usepackage{natbib}
\usepackage{url} 
\usepackage{float}
\usepackage{xcolor}
\usepackage{caption}
\usepackage{mathtools}
\usepackage{times}

\usepackage{bm}
\usepackage{pifont}
\usepackage{subfigure}
\usepackage[ruled,linesnumbered,lined,commentsnumbered]{algorithm2e}
\SetAlFnt{\small}
\SetAlCapFnt{\small}
\SetAlCapNameFnt{\small}
\usepackage{makecell}

\newcommand{\blind}{1}

\usepackage{authblk} 

\usepackage{threeparttable}
\usepackage{booktabs}
\usepackage{tabularx}
\usepackage{multirow}
\usepackage{multicol}
\usepackage{dcolumn}

\allowdisplaybreaks

\newtheorem{assumption}{Assumption}

\newtheorem{theory}{Theorem} 
\newtheorem{remark}{Remark}
\usepackage[colorlinks,
            linkcolor=black,
            anchorcolor=black,
            citecolor=black]{hyperref}

\begin{document}

\def\spacingset#1{\renewcommand{\baselinestretch}%
{#1}\small\normalsize} \spacingset{1}


\if1\blind
{
  \title{\bf Similarity-Informed Transfer Learning for Multivariate Functional Censored Quantile Regression}
  \author[a]{Hua Liu}
  \author[b]{Jiaqi Men}
  \author[b]{Shouxia Wang}
  \author[b]{Jinhong You$^*$}
  \author[c]{Jiguo Cao\thanks{Corresponding author: jiguo\_cao@sfu.ca ; johnyou07@163.com}}
  \affil[a]{\small School of Economics and Finance,
  Xi'an Jiaotong University, China}
  \affil[b]{\small School of Statistics and Management,
  Shanghai University of Finance and Economics, China}
  \affil[c]{\small Department of Statistics and Actuarial Science,
  Simon Fraser University, Canada}
  \date{}
  \maketitle
} \fi

\if0\blind
{
  \bigskip
  \bigskip
  \bigskip
  \begin{center}
    {\LARGE\bf Similarity-Informed Transfer Learning for Multivariate Functional Censored Quantile Regression}
\end{center}
  \medskip
} \fi

\bigskip
\begin{abstract}
To address the challenge of utilizing patient data from other organ transplant centers (source cohorts) to improve survival time estimation and inference for a target center (target cohort) with limited samples and strict data-sharing privacy constraints, we propose the \textbf{\texttt{S}}imilarity-\textbf{\texttt{I}}nformed \textbf{\texttt{T}}ransfer \textbf{\texttt{L}}earning (\texttt{SITL}) method. This approach estimates multivariate functional censored quantile regression by flexibly leveraging information from each source cohort based on its similarity to the target cohort. Furthermore, the method is adaptable to continuously updated real-time data. We establish the asymptotic properties of the estimators obtained using the \texttt{SITL} method, demonstrating improved convergence rates. Additionally, we develop an enhanced approach that combines the \texttt{SITL} method with a resampling technique to construct more accurate confidence intervals for functional coefficients, backed by theoretical guarantees. Extensive simulation studies and an application to kidney transplant data illustrate the significant advantages of the \texttt{SITL} method. Compared to methods that rely solely on the target cohort or indiscriminately pool data across source and target cohorts, the \texttt{SITL} method substantially improves both estimation and inference performance.

\end{abstract}

\noindent%

\spacingset{1.9} 
\newpage
\section{Introduction}\label{Intro}

Functional data refer to observations collected over multiple time points or other continuous domains \citep{horvath2012inference,wang2016functional,kokoszka2017introduction}, common in fields like biomedical research \citep{gao2024nonparametric}, economics, and finance. In the context of functional data analysis (FDA, \citealt{yao2005functional,lin2018mixture,
ye2020local,lee2020,li2021sparse,lee2023nonparametric,sang2024functional}), functional censored quantile regression \citep{jiang2020functional} has emerged as a more powerful tool for handling censored data, offering a distinct approach from the functional Cox model \citep{qu2016optimal}.
This model is particularly valuable in medical and biometric applications for exploring the relationship between a functional predictor and various quantiles of survival time. In organ transplant survival analysis, for instance, predicting patient survival based on indices observed during the follow-up period is crucial. Instead of focusing solely on mean survival times, clinicians are more concerned with how these functional predictors influence the high and low quantiles of survival times. This approach provides valuable insights into the distribution of survival times, helping doctors better understand the impact of follow-up indices on patient outcomes and informing clinical decisions.

More generally, to study the functional relationship between the multivariate functional predictors and the distribution of survival time in the presence of censoring, we introduce the multivariate functional censored quantile regression (mFCQR):
\begin{equation}\label{mFCQ}
    Q_{T}(\tau|\bm{X};\bm{\alpha}) = \exp\left\{ \int_{\mathcal{T}}
    \bm{X}^{\top}(s)\bm{\alpha}(s,\tau)\mathrm{d}s\right\},
\end{equation}
where $T$ denotes the survival time,
$\bm{X}(s)$ is the $p-$dimensional vector of functional predictors defined on the domain $\mathcal{T}$,
$Q_T(\tau|\bm{X};\bm{\alpha}) = \inf\{t:Pr(T\leq t|\bm{X})\geq \tau\}$ is the $\tau$th conditional quantile of $T$ given $\bm{X}$,
and $\bm{\alpha}(s,\tau)$ represents the covariate effect on the $\tau$th quantile of $T$, which is allowed to change with $\tau$. 
Notably, the proposed model encompasses several existing models as special cases.
For instance, when all components of $\bm{X}$ are invariant with respect to $s$, model (\ref{mFCQ}) reduces to the classical censored quantile regression model \citep{peng2008survival,zheng2018high,he2022scalable}. Besides, when only one component of $\bm{X}$ varies with $s$, 
it is simplified to the partially functional censored quantile regression model \citep{jiang2020functional}.

In practice, particularly with medical and biological data, samples from specific medical or organ transplant centers are often small, and censored data are commonly encountered, providing incomplete information. Additionally, factors such as geographic location may result in the target cohort representing only a specific subset of the population. When the sample size is small, the data from the target cohort may lack sufficient diversity and variability. In such scenarios, fitting a model solely using data from the target cohort for functional censored quantile regression (\ref{mFCQ}) with multiple infinite-dimensional functional predictors can lead to unstable estimates, high variance, and unreliable inferences.

A straightforward approach to address the above mentioned issues is to combine the target cohort (e.g. data from the target organ transplant center) with all source cohorts (e.g. data from other organ transplant centers) and fit the model (\ref{mFCQ}) using the pooled data. However, this method can lead to biased estimates due to the inherent differences between the target cohort and each source cohort, particularly when the source cohorts have much larger sample size. Another barrier of the pooled method is that sharing the raw data from the source cohorts directly is often infeasible due to privacy protection concerns in fields like medicine. This necessitates the exploration of alternative methods that are more effective and suitable for overcoming the challenges and difficulties mentioned above.

Alternatively, transfer learning has emerged as a widely recognized machine learning strategy for enhancing performance in a target cohort by leveraging information from distinct but related source cohorts \citep{zhuang2020comprehensive, bastani2021predicting}. This approach bridges the gap between target and source cohorts by adapting to the unique characteristics of the target cohort while utilizing the broader knowledge embedded in the source cohorts. In recent years, transfer learning has been extensively applied to statistical problems, primarily in the context of scalar data. Applications include high-dimensional linear models \citep{li2022transfer}, high-dimensional generalized linear models \citep{tian2023transfer,li2024estimation}, high-dimensional quantile regression \citep{bai2024transfer,zhang2022transfer}, nonparametric regression \citep{cai2024transfernon}, semiparametric regression models \citep{hu2023optimal,he2024representation}, and Cox models \citep{li2022transfer}, among others.

Despite its broad application in scalar data settings, the use of transfer learning in functional data analysis remains relatively limited. Notable contributions in this area include \cite{qin2024adaptive}, who investigated transfer learning for functional classification, \cite{cai2024transfer}, who developed a transfer learning method for functional mean estimation, and \cite{lin2022transfer}, who explored transfer learning in the context of functional linear regression models.

In this article, we address the challenges outlined in the third paragraph by proposing a similarity-informed transfer learning method, referred to as \texttt{SITL}, for estimating the multivariate functional censored quantile regression model (\ref{mFCQ}). Unlike most transfer learning applications, which require identifying an informative set of source cohorts and assigning equal weights to all detected sources, our \texttt{SITL} method takes a more nuanced approach. Specifically, \texttt{SITL} uses the loss function of the target cohort and each source cohort as a similarity measure across all quantile levels to determine the weight of each source. These similarity-based weights, combined with the sample sizes, are then used to compute a hybrid-weighted estimator from the source cohorts during the transfer step. Finally, the target cohort is utilized to refine and debias the estimator obtained in the transfer step, ensuring more accurate and reliable results.

The proposed \texttt{SITL} method has at least five advantages:
\begin{enumerate}
 \item 
The \texttt{SITL} method adaptively adjusts the importance of each source cohort by considering both its data similarity to the target cohort and its sample size.
    \item The \texttt{SITL} method is particularly well-suited for scenarios with data privacy concerns, as it enables the sharing of model estimators rather than individual-level data, thereby addressing privacy and feasibility issues.  
    \item The \texttt{SITL} is flexible in handling new datasets: when new source or target cohorts become available, the method can easily incorporate them without affecting previously established estimators, making it scalable and adaptable to evolving data environments.
    \item The \texttt{SITL} method improves the convergence rate from a theoretical perspective, leading to more efficient estimation of functional coefficients. This is attributed to the fact that the functional coefficients differing between the target cohort and all source cohorts, in contrast to the functional coefficients of the target cohort, are relatively simple and smooth.
    \item  When the \texttt{SITL} method is combined with resampling technique, it results in a smaller empirical variance of the estimators compared to the traditional resampling methods, thereby improving the reliability of statistical inference.
\end{enumerate}
These advantages collectively make \texttt{SITL} a powerful and practical approach transfer learning method for estimating the mFCQR model (\ref{mFCQ}).

The remainder of this article is organized as follows. Section \ref{Method} begins by introducing the baseline estimation for multivariate functional censored quantile regression, followed by the formulation of the transfer learning problem and the proposal of an algorithm. Section \ref{theory} presents the theoretical analysis of the proposed algorithms, develops a method for constructing confidence intervals, and establishes the asymptotic theory for these intervals. In Section \ref{simu}, we evaluate the numerical performance of our estimator through simulation studies. Section \ref{application} illustrates the application of our method with a real data analysis. Finally, Section \ref{conclu} offers conclusions and discusses potential future directions.

\section{Similarity-Informed Transfer Learning}\label{Method}

\subsection{Baseline Estimation}
\label{Baseline}
Let $C$ denote the censoring time, $Y = T\wedge C$ and $\delta = I(T\leq C)$, where $\wedge$ is the minimum operator and $I(\cdot)$ represents the indicator function.  We assume the censoring time $C$ is independent of $T$ conditional on multiple functional predictors. 
Define the conditional cumulative distribution function (CDF) of $T$ as $F_T(t|\bm{X})=\mathrm{Pr}(T\leq t|\bm{X})$, and 
the cumulative hazard function as $\Lambda_T(t|,\bm{X}) = -\log\{1-F_T(t|\bm{X})\}$. Let the counting process be $N(t) = I(Y\leq t, \delta=1)$. Then, the process 
$M(t)=N(t)-\Lambda_T(t \wedge Y |\bm{X})$ forms a martingale. Consequently,
using the property of martingale, we have $\mathbb{E}(M(t)|\bm{X})=0$.

Suppose that the full sample consists of $n$ independent and identically distributed (i.i.d.) observations of $(Y,\delta, \bm{X}(t), t\in\mathcal{T})$, denoted by $\left\{(Y_i,\delta_i, \bm{X}_i(t), t\in\mathcal{T}),i=1,\ldots,n\right\}$. 
For each individual sample, let $N_i(t)$ and $M_i(t)$ represent the sample versions of $N(t)$ and $M(t)$, respectively, and $\Lambda_T(t|\bm{X}_i)$ is the cumulative hazard function of
$T_i$ conditional on $\bm{X}_i$. 
According to \cite{peng2008survival} and \cite{jiang2020functional}, $M_i(t)$ satisfies
$\mathbb{E}\left\{M_i(t)|\bm{X}_i\right\}=0$ for $t\geq 0$ and $i=1,...,n$. For any given $\tau$, we have
\begin{equation}\label{martmean}
    \mathbb{E}\left\{M_i\left[ Q_{T}(\tau|\bm{X}_i;\bm{\alpha}_0)\right]\right\}=0,
    \quad \text{for} \quad i=1,...,n,
\end{equation}
where $\bm{\alpha}_0(\cdot,\tau)$ are the true coefficient functions. 

In the martingale $M_i(t)$, we have $\Lambda_T\left\{\left[ Q_{T}(\tau|\bm{X}_i;\bm{\alpha}_0)\right]\wedge Y_i|\bm{X}_i\right\}
     = \int_{0}^{\tau} I\left\{Y_i\geq Q_{T}(u|\bm{x}_i;\bm{\alpha}_0)\right\} \\\mathrm{d} H(u)$,
where $H(u) = -\log(1-u),\ 0\leq u<1$. To compute this integration, define $\mathcal{A}_{\tau,L}=\{0=\tau_0<\tau_1<\cdots<\tau_L=\tau_{U}<1\}$. For $\tau_j,\ j>0$, we obtain 
\begin{align}
      \hat{\Lambda}_T\left\{\left[ Q_{T}(\tau|\bm{X}_i;\bm{\alpha}_0)\right]\wedge Y_i|\bm{X}_i\right\}
     =\sum_{l=0}^{j-1} I\left\{Y_i\geq Q_{T}(\tau_l|\bm{X}_i;\bm{\alpha}_0)\right\} \times  \left\{H(\tau_{l+1})-H(\tau_l)\right\}.
     \label{appcumhazard}
\end{align}
In addition, $Q_{T}(\tau_0|\bm{X}_i;\bm{\alpha}_0)=0$.
For the coefficient functions $ \alpha_{d}(s,\tau),d=1,...,q$, we approximate them using B-spline basis functions \citep{de_Boor_2001}. Suppose there are $M_d$ equally spaced interior knots in the domain $\mathcal{T}$, and let $B_{dl}(s),l=1,\ldots,M_d+r$ represent the corresponding $r$-th order B-spline basis functions. Then, we express $ \alpha_{d}(s,\tau)$ as $\alpha_{d}(s,\tau) \approx \sum_{l=1}^{M_d+r}B_{dl}(s)\gamma_{dl}(\tau)=\bm{B}^{\top}_{d}(s)\bm{\gamma}_d(\tau)$,
where
$\bm{B}_d(s) = (B_{d1}(s),\cdots,B_{d(M_d+r)}(s))^{\top}$ and 
$\bm{\gamma}_d(\tau) = (\gamma_{d1}(\tau),\cdots,\gamma_{d(M_d+r)}(\tau) )^{\top}$. 
Let $\int_{\mathcal{T}}
    X_{id}(s)\bm{B}_{d}(s)\mathrm{d}s = \langle X_{id},\bm{B}_{d}\rangle \overset{\Delta}{=}\bm{W}_{id}$, $\bm{W}_i=(\bm{W}_{i1}^{\top},\cdots,\bm{W}_{iq}^{\top})^{\top}$ and $\bm{\gamma}(\tau) = (\bm{\gamma}_1^{\top}(\tau),\cdots,
    \bm{\gamma}_q^{\top}(\tau))^{\top}$.
Thus, the mFCQR model (\ref{mFCQ}) can be expressed as 
\begin{align}
    Q_{T}(\tau|\bm{X}_i;\bm{\alpha}) 
    =  \exp\left\{ \sum_{d=1}^q\int_{\mathcal{T}}
    X_{id}(s)\alpha_d(s,\tau)\mathrm{d}s\right\}
    \approx 
     \exp\left\{ \bm{W}_i^{\top} \bm{\gamma}(\tau)\right\}. \label{mFCQapprox}
\end{align}

By combining the approximation of cumulative hazard function (\ref{appcumhazard}) and the mFCQR model approximation
(\ref{mFCQapprox}), based on the 
mean zero property of martingale (\ref{martmean}),
the estimators $\hat{\bm{\gamma}}(\tau_j)$ for $j=1,...,L$ can be obtained by sequentially solving the following estimating equation:
\begin{align}
      \bm{0} = & \sum_{i=1}^n \bm{W}_i
      \Bigg\{N_i\left[Q_{T}(\tau_j|\bm{W}_i;\bm{\gamma})\right]  - \sum_{l=0}^{j-1} I\left\{Y_i\geq Q_{T}(\tau_l|\bm{W}_i;\hat{\bm{\gamma}}(\tau_l))\right\} \times  \left\{H(\tau_{l+1})-H(\tau_l)\right\}
      \Bigg\},\label{estequation}
\end{align}
where we set $Q_{T}(\tau_0|\bm{W}_i;\hat{\bm{\gamma}}(\tau_0))=0$ for $i=1,...,n$. The estimating equation (\ref{estequation}) is the gradient of a convex function. Thus, solving (\ref{estequation}) is equivalent to minimizing the following function with respect to $\bm{\gamma}(\tau_j)$:
\begin{align*}
      - & \sum_{i=1}^n \Big[\log(Y_i)-\bm{W}_i^{\top} \bm{\gamma}(\tau_j)\Big] \nonumber
     \times \Bigg\{N_i\left[Q_{T}(\tau_j|\bm{W}_i;\bm{\gamma}(\tau_j))\right]\nonumber\\    
     & \qquad \qquad - \sum_{l=0}^{j-1} I\left\{Y_i\geq Q_{T}(\tau_l|\bm{W}_i;\hat{\bm{\gamma}}(\tau_l))\right\} \times  \left\{H(\tau_{l+1})-H(\tau_l)\right\}
      \Bigg\}.
\end{align*}
By extending Theorem 2 of \cite{jiang2020functional}, the estimator 
$\hat{\alpha}_d(s,\tau)= \bm{B}^{\top}_{d}(s)\hat{\bm{\gamma}}_d(\tau)$ attains the following optimal convergence rate $\sup_{\substack{\tau\in[0,\tau_{U}]
    }}\|\hat{\alpha}_d(\cdot,\tau)-\alpha_{d}(\cdot,\tau)\|_{L_2} = O_p((M_{d}/n)^{1/2}+M_{d}^{-\theta})$,
where $\theta$ represents the smoothness of the function $\alpha_{d}(\cdot,\cdot)$, specially the number of its derivatives, and $\|\cdot\|_{L_2}$ to denote the $L_2$ norm. 
The optimal nonparametric convergence rate is achieved when $M=O(n^{1/(2\theta+1)})$, yielding a rate of $n^{-\theta/(2\theta+1)}$.  However, when the functional coefficients are less smooth, meaning the functions are rougher and more complex, the convergence rate becomes slower due to the increased difficulty in estimating such functions.

\subsection{Similarity-Informed Transfer Learning}\label{transfermethod}
Suppose the target data is given as
$\{\bm{X}^{(0)}_i(t),T^{(0)}_i;
t\in \mathcal{T}^{(0)}\}_{i=1}^{n_0}$ with functional coefficients $\bm{\alpha}^{(0)}(s,\tau)$. Additionally, we have data from $K$ source cohorts, where the $k$th source is denoted by  $\{\bm{X}^{(k)}_i(t),T^{(k)}_i;t\in \mathcal{T}^{(k)} k\leq K$, with potentially different functional coefficients denoted by $\bm{\alpha}^{(\mathcal{S},k)}(s,\tau)$. The difference in functional coefficients between the $k$-th source cohort and the target cohort is defined as 
$\bm{\alpha}^{(\eta,k)}(s,\tau)=\bm{\alpha}^{(0)}(s,\tau)-\bm{\alpha}^{(\mathcal{S},k)}(s,\tau)$.
In practice, the similarity of each source cohort to the target cohort can vary widely. Source cohorts with very few observations or those significantly different from the target cohort may contribute little to the analysis. Furthermore, including too many irrelevant sources in transfer learning can degrade the performance on the target cohort. 

\begin{figure}[htbp]
    \centering
    \includegraphics[width=16cm]{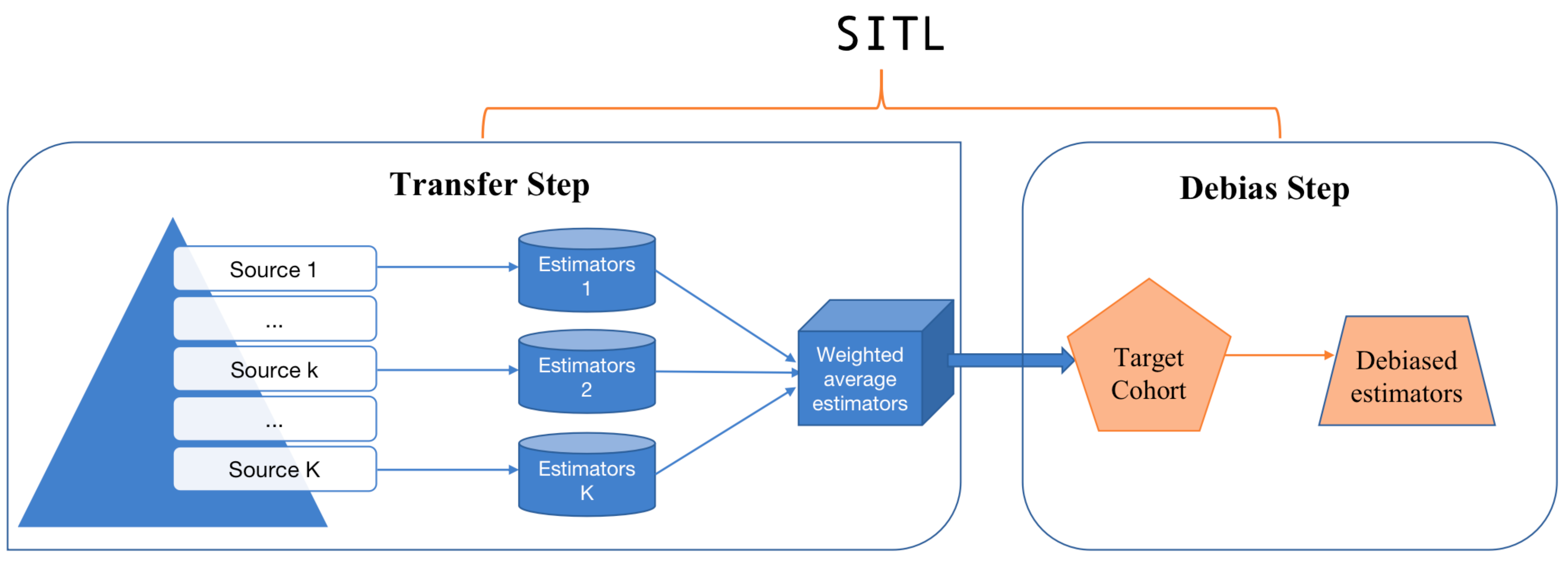}
    \caption{The schematic diagram for the Similarity-Informed Transfer Learning (\texttt{SITL}) method.}
    \label{flowchart}
\end{figure}

To effectively leverage information from informative source cohorts while minimizing the impact of irrelevant ones, we propose the  \texttt{SITL} algorithm, in this section. 
In this method, we incorporate information from all source cohorts based on their similarity to the target cohort and the size of their samples. The flowchart of \texttt{SITL} is presented in Figure \ref{flowchart}. The subsequent two subsections provide detailed explanations of the steps for similarity weight calculation, transfer and debias.

\subsubsection{Calculating Similarity Weights}
\label{Similaritycal}
We begin by discussing how to measure the similarity between the $k$-th source cohort to the target cohort.
First, we randomly divide the target cohort into two equal-size subsets: $\mathcal{I}$ and $\mathcal{I}^c$. 
For the target cohort, we approximate the functional coefficient $\alpha_{d}^{(0)}(s,\tau)$ as $\alpha_{d}^{(0)}(s,\tau) \approx \sum_{l=1}^{M_{d}+r}B_{dl}(s)\gamma_{dl}^{(0)}(\tau)=\bm{B}^{(0)\top}_{d}(s)\bm{\gamma}^{(0)}_d(\tau)$,
where
$\bm{B}_d^{(0)}(s) = (B_{d1}(s),\cdots,B_{d(M_d+r)}(s))^{\top}$ and 
$\bm{\gamma}^{(0)}_d(\tau)= \left(\gamma_{d1}(\tau),\cdots,\gamma_{d(M_d+r)}(\tau) \right)^{\top}$. 
Let $\langle X^{(0)}_{id},\bm{B}^{(0)}_{d}\rangle \overset{\Delta}{=}\bm{W}^{(0)}_{id}$, $\bm{W}^{(0)}_i=(\bm{W}_{i1}^{(0)\top},\cdots,\bm{W}_{iq}^{(0)\top})^{\top}$ and $\bm{\gamma}^{(0)}(\tau) = (\bm{\gamma}_1^{(0)\top}(\tau),\cdots,\bm{\gamma}_q^{(0)\top}(\tau))^{\top}$. 
Using the baseline estimation as stated in Section \ref{Baseline}, the estimators $\hat{\bm{\gamma}}^{(0)}(\tau)$ using the $\mathcal{I}$ are obtained by solving the following equation with respect to $\bm{\gamma}^{(0)}(\tau_j)$:
\begin{align*}
      \bm{0} = & \sum_{i\in \mathcal{I}}\bm{W}^{(0)}_i
      \Bigg\{N_i\left[Q_{T}(\tau_j|\bm{W}^{(0)}_i;\bm{\gamma}^{(0)})\right]\nonumber \\
      & \qquad \qquad \qquad - \sum_{l=0}^{j-1} I\left\{Y_i\geq Q_{T}(\tau_l|\bm{W}^{(0)}_i;\hat{\bm{\gamma}}^{(0)}(\tau_l))\right\} \times  \left\{H(\tau_{l+1})-H(\tau_l)\right\}
      \Bigg\},
\end{align*}
and the corresponding estimators are given by $
\hat{\alpha}_{d}^{(0)}(s,\tau) =\bm{B}^{(0)\top}_{d}(s)\hat{\bm{\gamma}}^{(0)}_d(\tau)$.

Similarly, for the $k$-th source cohort, we define 
$\bm{B}_d^{(k)}(s) = (B_{d1}(s),\cdots,B_{d(M_{d,k}+r)}(s))^{\top}$,
$\bm{\gamma}^{(k)}_d(\tau) = (\gamma_{d1}(\tau),\cdots,\gamma_{d(M_{d,k}+r)}(\tau) )^{\top}$,
$\langle X^{(k)}_{id},\bm{B}^{(k)}_{d}\rangle \overset{\Delta}{=}\bm{W}^{(k)}_{id}$, $\bm{W}^{(k)}_i=(\bm{W}_{i1}^{(k)\top},\cdots,\bm{W}_{iq}^{(k)\top})^{\top}$ and $\bm{\gamma}^{(k)}(\tau) = (\bm{\gamma}_1^{(k)\top}(\tau),\cdots,\bm{\gamma}_q^{(k)\top}(\tau))^{\top}$. The estimators for the $k$th source cohort are obtained by solving the following equation with respect to $\bm{\gamma}^{(k)}(\tau_j)$:
\begin{align*}
      \bm{0} = & \sum_{i=1}^n \bm{W}_i^{(k)}
      \Bigg\{N_i\left[Q_{T}(\tau_j|\bm{W}_i^{(k)};\bm{\gamma}^{(k)})\right]\nonumber \\
     & \qquad \qquad \qquad - \sum_{l=0}^{j-1} I\left\{Y_i\geq Q_{T}(\tau_l|\bm{W}_i^{(k)};\hat{\bm{\gamma}}^{(k)}(\tau_l))\right\} \times  \left\{H(\tau_{l+1})-H(\tau_l)\right\}
      \Bigg\},
\end{align*}
and the corresponding estimators are $
\hat{\alpha}_{d}^{(\mathcal{S},k)}(s,\tau) =\bm{B}^{(k)\top}_{d}(s)\hat{\bm{\gamma}}^{(k)}_d(\tau)$.

To assess the performance of each source cohort, we calculate the empirical loss function for the subset $\mathcal{I}^c$ using the estimators obtained from each source cohort and the subset $\mathcal{I}$ of the target cohort:
\begin{align}\label{loss_tar}
    \hat{\mathcal{L}}(\hat{\bm{\alpha}}^{(\mathcal{S},k)}(\tau_j)|\mathcal{I}^c) = - & \sum_{i\in \mathcal{I}^c} \Big[\log(Y_i)-\langle \bm{X}_i^{(0)\top},\hat{\bm{\alpha}}^{(\mathcal{S},k)}(\tau_j)\rangle\Big] 
     \times \Bigg\{N_i\left[Q_{T}(\tau_j|\bm{X}_i^{(0)\top},\hat{\bm{\alpha}}^{(\mathcal{S},k)}(\tau_j)\right]\nonumber\\
     & \qquad \qquad - \sum_{l=0}^{j-1} I\left\{Y_i\geq Q_{T}(\tau_l|\bm{X}_i^{(0)};\hat{\bm{\alpha}}^{(\mathcal{S},k)}(\tau_l))\right\} \times  \left\{H(\tau_{l+1})-H(\tau_l)\right\}
      \Bigg\}, 
\end{align}
and we can get the loss on the subset $\mathcal{I}$ of the target cohort $\hat{\mathcal{L}}(\hat{\bm{\alpha}}^{(0)}(\tau_j)|\mathcal{I}^c)$ by replacing $\hat{\bm{\alpha}}^{(\mathcal{S},k)}(\tau_j)$ in $\hat{\mathcal{L}}(\hat{\bm{\alpha}}^{(\mathcal{S},k)}(\tau_j)|\mathcal{I}^c)$ with $\hat{\bm{\alpha}}^{(0)}(\tau_j)$.
As indicated by the sequential nature of the estimation procedure in Section \ref{Baseline}, we compute the difference between the two loss functions,  
$\hat{\mathcal{L}}(\hat{\bm{\alpha}}^{(\mathcal{S},k)}(\tau_j)|\mathcal{I}^c)$ and $\hat{\mathcal{L}}(\hat{\bm{\alpha}}^{(0)}(\tau_j)|\mathcal{I}^c)$ across all quantile levels to calculate the similarity weights $\hat{\omega}_k,k=1,...,K$:
\begin{equation}
        \label{diff}
        \hat{\omega}_k=\frac{1}{h}
       \mathcal{K}(\frac{\hat{\mathcal{D}}^{(k)}}{h}),\quad \hat{\mathcal{D}}^{(k)} = \frac{1}{L}\sum_{j=1}^{L} 
        \left\{
        \hat{\mathcal{L}}(\hat{\bm{\alpha}}^{(\mathcal{S},k)}(\tau_j)|\mathcal{I}^c)-
        \hat{\mathcal{L}}(\hat{\bm{\alpha}}^{(0)}(\tau_j)|\mathcal{I}^c)
        \right\},
    \end{equation}
where $\mathcal{K}(u)= (2\pi)^{-1/2}e^{-u^2/2}$ 
is the Gaussian kernel function and $h$ is bandwidth.
By adjusting the bandwidth $h$, we can control the influence of each source cohort. Specifically, a larger $h$ assigns smaller weights to a broader range of sources, while a smaller $h$ focuses on fewer sources, assigning them larger weights.

\subsubsection{Transfer Step}
\label{transdebias}
In the transfer step, we incorporate information from all source cohorts by weighting their estimators using the similarity weights $\hat{\omega}_k$ and sample sizes $n_k$, as given by:
\begin{equation*}\label{transtep}
                    \begin{split}
                        \hat{\bm{\alpha}}^{(\mathcal{S})}(s,\tau) & =  \frac{1}{\sum_{k=1}^K n_k\hat{\omega}_k} \sum_{k=1}^K n_k \hat{\omega}_k \hat{\bm{\alpha}}^{(\mathcal{S},k)}(s,\tau).
                    \end{split}
                    \end{equation*}
Denote 
$\bm{\alpha}^{(\eta)}(s,\tau)=\bm{\alpha}^{(0)}(s,\tau)-\bm{\alpha}^{(\mathcal{S})}(s,\tau)$ represents the overall difference in functional coefficients between the target cohort and all source cohorts, capturing the unique characteristics of the target cohort.
When $\bm{\alpha}^{(\eta)}(s,\tau)\equiv 0$ for all $s\in \mathcal{I}$ and $\tau\in(0,\tau_U]$, 
all source cohorts are identical to the target cohort in terms of mFCQR.
In practice, differences between the target and source cohorts are common, leading to biased estimators in the transfer step.

\subsubsection{Debias Step}
In the Debias Step, we aim to estimate $\bm{\alpha}^{(\eta)}(s,\tau)$ based on the target cohort in order to correct or debias the estimators from the transfer step.
We approximate $\bm{\alpha}^{(\eta)}(s,\tau)$
as $\alpha_d^{(\eta)}(s,\tau)=\sum_{l=1}^{M_d^{\eta}+r} B_{dl}(s)\gamma_{dl}^{(\eta)}(\tau)=\bm{B}_d^{(\eta)\top}(s)\bm{\gamma}_d^{(\eta)}(\tau)$,
where 
$\bm{B}_d^{(\eta)}(s) = (B_{d1}(s),\cdots,B_{d(M^{\eta}_d+r)}(s))^{\top}$ and 
$\bm{\gamma}^{(\eta)}_d(\tau)= (\gamma_{d1}^{(\eta)}(\tau),\cdots,\gamma^{(\eta)}_{d(M_d^{(\eta)}+r)}(\tau) )^{\top}$. 
Let $\langle X^{(0)}_{id},\bm{B}^{(\eta)}_{d}\rangle \overset{\Delta}{=}\bm{W}^{(\eta)}_{id}$, $\bm{W}^{(\eta)}_i=(\bm{W}_{i1}^{(\eta)\top},\cdots,\bm{W}_{iq}^{(\eta)\top})^{\top}$ and $\bm{\gamma}^{(\eta)}(\tau) = (\bm{\gamma}_1^{(\eta)\top}(\tau),\cdots,\bm{\gamma}_q^{(\eta)\top}(\tau))^{\top}$. 
Define the $\tau_j$-th quantile residual as
$\hat{e}_i^{(\mathcal{S})}(\tau_j)=\log(Y_i)-
\langle \bm{X}_i^{\top}, \hat{\bm{\alpha}}^{(\mathcal{S})}(\tau_j)\rangle$,
then for any given quantile level $\tau_j$, 
the objective function in the Debias Step can be rewritten as:
\begin{align}
      - & \sum_{i=1}^n \left\{\hat{e}_i^{(\mathcal{S})}(\tau_j)-\bm{W}_i^{(\eta)\top} \bm{\gamma}^{(\eta)}(\tau_j)\right\}  \times 
     \Bigg\{ I \left[
    \hat{e}_i^{(\mathcal{S})}(\tau_j)-\bm{W}_i^{(\eta)\top} \bm{\gamma}^{(\eta)}(\tau_j)\leq 0; \delta_i=1
     \right]-u_{ij}
     \Bigg\}\nonumber,\\
     \label{debais2}
\end{align}
with $u_{ij}=\sum_{l=0}^{j-1} I\left\{\hat{e}_i^{(\mathcal{S})}(\tau_l)\geq \bm{W}_i^{(\eta)\top} \hat{\bm{\gamma}}^{(\eta)}(\tau_l) \right\} \times  \left\{H(\tau_{l+1})-H(\tau_l)\right\}$
based on the debias estimators under $\tau_l,l=0,...,j-1$.
The gradient of the objective function (\ref{debais2}) with respect to 
$\bm{\gamma}^{(\eta)}$ is 
\begin{align}
       & \sum_{i=1}^n \bm{W}_i^{(\eta)}
        \Bigg\{ I \left[
    \hat{e}_i^{(\mathcal{S})}(\tau_j)-\bm{W}_i^{(\eta)\top} \bm{\gamma}^{(\eta)}(\tau_j)\leq 0; \delta_i=1
     \right]-u_{ij}
     \Bigg\}.
     \label{gradient}
\end{align}
To solve (\ref{gradient}), we follow these steps:
\begin{itemize}
    \item \textbf{Step 0}: Initialize $\hat{e}_i^{(\mathcal{S})}(\tau_0)=0$ for all $i=1,2,...,n_0$, and $\hat{\bm{\alpha}}^{(\eta)}(s,\tau_0)=0$.
    \item  \textbf{Step I}: Given $\tau_j,j>1$, we need to calculate $\hat{e}_i^{(\mathcal{S})}(\tau_j)$ and
    $u_{ij}$ for all $i=1,2,...,n_0$.
    \item  \textbf{Step II}: Use an iterative algorithm to minimize the objective function in Equation (\ref{debais2}) with the gradient given in Equation (\ref{gradient}) to obtain the estimators $\hat{\bm{\alpha}}^{(\eta)}(s,\tau_j)=\bm{B}_d^{(\eta)\top}(s)\hat{\bm{\gamma}}_d^{(\eta)}(\tau_j)$.
    \item Repeat \textbf{Step I} and \textbf{Step II} until we get the estimators for the last quantile level $\tau_U$.
\end{itemize}

Finally, the estimators of the functional coefficients for the target cohort, obtained using the \texttt{SITL} algorithm after the calculation of similarity weight, the transfer step and the debias step, are given by
\begin{align*}
    \label{SITL}
    \tilde{\bm{\alpha}}(s,\tau)= \hat{\bm{\alpha}}^{(\mathcal{S})}(s,\tau)+\hat{\bm{\alpha}}^{(\eta)}(s,\tau).
\end{align*}
To provide readers with a clearer understanding of the entire \texttt{SITL} algorithm, we formally present the detailed steps in Algorithm \ref{altrans3}.

\begin{algorithm}[htbp]
  \footnotesize
\SetKwData{Left}{left}\SetKwData{This}{this}\SetKwData{Up}{up}
\SetKwFunction{Union}{Union}\SetKwFunction{FindCompress}{FindCompress}
\SetKwInOut{Input}{Input}\SetKwInOut{Output}{Output}
\Input{Target data, all source data, and quantile levels}
\BlankLine
\begin{itemize}
			\item \textbf{Similarity Weights Calculation}: 
               \begin{itemize}
    \item \textbf{Step 1:} Randomly split the target cohort into two equal-size subsets with the same sample size $n_0/2$: $\mathcal{I}$ and $\mathcal{I}^c$, and $\mathcal{I}\cup \mathcal{I}^c = \{1,2,...,n_0\} $. 
    \item \textbf{Step 2:} Calculate 
$\hat{\bm{\alpha}}^{(0)}(s,\tau)$ based on the subset $\mathcal{I}$.
    \item \textbf{Step 3:} For each source cohort $1\leq k\leq K$, we calculate 
    $\hat{\bm{\alpha}}^{(\mathcal{S},k)}(s,\tau)$ based on the $k$-th source.
    \item \textbf{Step 4:} For each $1\leq k\leq K$, we compute the difference between the loss functions $\hat{\mathcal{L}}(\hat{\bm{\alpha}}^{(0)}(\tau)|\mathcal{I}^c)$ and $\hat{\mathcal{L}}(\hat{\bm{\alpha}}^{(\mathcal{S},k)}(\tau)|\mathcal{I}^c)$ on the subset $\mathcal{I}^c$:
    \begin{equation*}
        \hat{\mathcal{D}}^{(k)} = \frac{1}{L}\sum_{j=1}^{L} 
        \left\{
        \hat{\mathcal{L}}(\hat{\bm{\alpha}}^{(\mathcal{S},k)}(\tau_j)|\mathcal{I}^c)-
        \hat{\mathcal{L}}(\hat{\bm{\alpha}}^{(0)}(\tau_j)|\mathcal{I}^c)
        \right\}.
    \end{equation*}
    \item \textbf{Step 5:} The weight of source cohorts is $\hat{\omega}_k=1/h
       \mathcal{K}(\hat{\mathcal{D}}^{(k)}/h)$, where $\mathcal{K}(u)= (2\pi)^{-1/2}e^{-u^2/2}$ is 
       Gaussian kernel function and $h$ is bandwidth.
       \end{itemize}
       			\item \textbf{Transfer Step}: 
                    Take a kernel-based weighted average of $\hat{\bm{\alpha}}^{(\mathcal{S},k)}(s,\tau)$ and denote by $\hat{\bm{\alpha}}^{(\mathcal{S})}(s,\tau)$:
                    \begin{equation*}
                    \begin{split}
                    \hat{\bm{\alpha}}^{(\mathcal{S})}(s,\tau) & =  \frac{1}{\sum_{k=1}^K n_k\hat{\omega}_k} \sum_{k=1}^K n_k \hat{\omega}_k \hat{\bm{\alpha}}^{(\mathcal{S},k)}(s,\tau).
                    \end{split}
                    \end{equation*}
                    
			\item \textbf{Debias Step:} 
            Let  the differences 
 $\alpha_d^{(\eta)}(s,\tau)=\sum_{l=1}^{M_d^{\eta}+r} B_{dl}(s)\gamma_{dl}^{(\eta)}(\tau)=\bm{B}_d^{(\eta)\top}(s)\bm{\gamma}^{(\eta)}(\tau),$
            then solve the following estimation equation with respect to $\bm{\gamma}(\tau_j)$ at given quantile level $\tau_j$:
    \begin{align*}
       & \sum_{i=1}^n \bm{W}_i^{(\eta)}
        \Bigg\{ I \left[
    \hat{e}_i^{(\mathcal{S})}(\tau_j)-\bm{W}_i^{(\eta)\top} \bm{\gamma}^{(\eta)}(\tau_j)\leq 0; \delta_i=1
     \right]-u_{ij}
     \Bigg\},
\end{align*}
with $u_{ij}=\sum_{l=0}^{j-1} I\left\{\hat{e}_i^{(\mathcal{S})}(\tau_l)\geq \bm{W}_i^{(\eta)\top} \hat{\bm{\gamma}}^{(\eta)}(\tau_l) \right\} \times  \left\{H(\tau_{l+1})-H(\tau_l)\right\}$.
\end{itemize}
\BlankLine
\Output{
$\tilde{\bm{\alpha}}(s,\tau)= \hat{\bm{\alpha}}^{(\mathcal{S})}(s,\tau)+\hat{\bm{\alpha}}^{(\eta)}(s,\tau)$.
}
\caption{
The
Similarity-Informed Transfer Learning Method for Estimating the Multivariate Functional Censored Quantile Regression Model (\ref{mFCQ})
}\label{altrans3}
\end{algorithm}

The steps of the \texttt{SITL} algorithm
show that the weight calculation and transfer steps do not require the combination of all source cohorts. Therefore,
we can share the estimators from the source cohorts, rather than the individual-level data, which is beneficial when data-sharing policies are constrained by privacy or other concerns \citep{litransfer}.
Additionally, the \texttt{SITL} algorithm is flexible when new source cohorts or target cohorts are introduced. 
When a new source cohort comes, we only need to calculate its similarity weight, update the weighted estimators in the transfer step, and resolve the debias objective function. This procedure does not affect the estimators from the existing source cohorts.
Similarly, when a new target cohort emerges, we can easily re-calculate the similarity weights by directly using the estimators already available from the source cohorts.

Another interesting feature of the \texttt{SITL} algorithm is the ability to adjust the importance of each source cohort by selecting a different kernel function.
For example, if we choose a uniform kernel function, $\mathcal{K}(u) = 0.5, |u|\leq 1$, all informative sources receive the same weight $\hat{\omega}_i = 0.5\times I\left\{|\hat{\mathcal{D}}^{(k)}|
       \leq h\right\}$. 
In this case, the Algorithm \ref{altrans3}
reduces to a hard-threshold transfer learning algorithm, called \texttt{Trans\_HT}, which is commonly used in existing applications of transfer learning for high dimensional linear, generalized, quantile regression and functional regressions. 
For further details on the \texttt{Trans\_HT} algorithm, please refer to Section S1
of the supplementary document. In the simulation and application sections, we demonstrate that the proposed \texttt{SITL} method outperforms the \texttt{Trans\_HT} method by flexibly assigning different weights to sources based on their similarity to the target cohort. For more details, please refer to Section \ref{simu} and Section \ref{application}.

\section{Theoretical Results}
\label{theory}
We first define some notations. The true values of $\alpha_d(s,\tau)$ in the target cohort are denoted by
$\alpha_{d}^{(0)}(s,\tau)$. 
We use $\|\cdot\|_{L_2}$ to denote the $L_2$ norm and use $\|\cdot\|_{\infty}$ to denote the sup norm.
For an arbitrary vector $\bm{a}$, let $\bm{a}^{\otimes 2}=\bm{a}\bm{a}^{\top}$. Denote the space of $r$-order smooth functions defined on $\mathcal{T}$ as $\mathcal{C}^{(r)}(\mathcal{T})=\{m|m^r\in \mathcal{C}(\mathcal{T})\}$, where $\mathcal{C} (\mathcal{T})$ is the collection of real-valued functions that are bounded and continuous in $\mathcal{T}$.
Further, we define
$F_T(t|\bm{X})=\mathrm{Pr}(T\leq t|\bm{X})$,
   $f(t|\bm{X}) = d F_T(t|\bm{X})/dt$,
   $\bar{F}_T(t|\bm{X})=1-F_T(t|\bm{X})$, 
   $\bar{f}(t|\bm{X}) = d \bar{F}_T(t|\bm{X})/dt=-d F_T(t|\bm{X})/dt=-f(t|\bm{X})$, 
    $\tilde{F}_T(t|\bm{X})=\mathrm{Pr}(T\leq t,\delta=1|\bm{X})$ and 
   $\tilde{f}(t|\bm{X}) = d \tilde{F}_T(t|\bm{X})/dt$.
Similar with \cite{peng2008survival} and \cite{jiang2020functional}, define $\mathcal{B}^{(k)}(\vartheta_k,\sum_{d=1}^q M_{d,k}+qr) = \{\bm{b}\in \mathbb{R}^{\sum_{d=1}^q M_{d,k}+qr}: \inf_{\tau\in(0,\tau_U]}\|  \mathbb{E}\{ \bm{W}_i^{(k)}
          N [
\exp(\bm{W}_i^{(k)\top} \bm{b}(\tau))
     ]\}- \mathbb{E}\{ \bm{W}_i^{(k)}
          N [
\exp(\bm{W}_i^{(k)\top} \bm{\gamma}^{(k)}(\tau))
     ]\}\|_{\infty}
     \leq \vartheta_k\}$ for $k=0,1,2...,K$ and 
$\mathcal{B}^{(\eta)}(\vartheta_{\eta},\sum_{d=1}^q M^{(\eta)}_{d}+qr) = \{\bm{b}\in \mathbb{R}^{\sum_{d=1}^q M^{(\eta)}_{d}+qr}: \inf_{\tau\in(0,\tau_U]}\|  \bm{\mu}\left(\bm{b}|\hat{\bm{\alpha}}^{\mathcal{(S)}} \right) 
- \bm{\mu}\left(\bm{\gamma}^{(\eta)}(\tau)|\hat{\bm{\alpha}}^{\mathcal{(S)}} \right) \|_{\infty}\leq \vartheta_{\eta}\}$.
In addition, $\mathcal{B}^{(k)}(\vartheta_{0,k},\sum_{d=1}^q M_{d,k}+qr)$ is a neighborhood that contains $\{\bm{\gamma}^{(k)},\tau\in(0,\tau_U]\}$ and $\mathcal{B}^{(\eta)}(\vartheta_{0,\eta},\sum_{d=1}^q M^{(\eta)}_{d}+qr)$ is a neighborhood that contains $\{\bm{\gamma}^{(\eta)},\tau\in(0,\tau_U]\}$.

Let $\phi$ be a map $\phi\left[ \bm{A}(\tau),\bm{B}(\tau),\bm{C}(\tau) \right]=\int_0^{\tau}
    \prod_{u\in(s,\tau]}
    \left[\bm{I}+\bm{B}(u)\bm{C}(u)^{-1}dH(u)\right]d\bm{A}(s)$
for arbitrary matrices valued functions $\bm{A}$, $\bm{B}$ and $\bm{C}$, with the product integral $\prod$ defined by \cite{gill1990survey}.
We also define $\bm{\iota}_i(\tau)=\bm{W}_i^{(\eta)}
         \{ N [
\exp(\langle \bm{X}_i^{(0)},\bm{\alpha}^{(0)}(\tau)\rangle
     ] - \int_0^{\tau}
     I [Y_i\geq
\exp(\langle \bm{X}_i^{(0)},
\bm{\alpha}^{(0)}(u)\rangle
     ]d H(u)\}$, 
the corresponding Hessian matrix $\bm{\mathcal{J}}(\bm{\alpha}_0(\tau)) =\mathbb{E}\{ (\bm{W}_i^{(\eta)})^{\otimes 2}
          \tilde{f} [
\exp(\langle \bm{X}_i^{(0)},
\bm{\alpha}^{(0)}(\tau)\rangle)
     ]\times \exp(\langle \bm{X}_i^{(0)},\bm{\alpha}^{(0)}(\tau)\rangle)\}$,
and the variance and covariance components 
$\tilde{\bm{\mathcal{J}}}(\bm{\alpha}^{(0)}(\tau))$ with replacing
$\tilde{f}(\cdot)$ in $\bm{\mathcal{J}}(\bm{\alpha}_0(\tau))$ with 
$\bar{f}(\cdot)$.

\subsection{Assumptions}
\label{assumps}
To establish the asymptotic results of the estimated functional coefficients, the following assumptions are needed.

\begin{assumption}\label{assumpofmoment}
\rm{For the functional predictors $X_d^{(k)}(s);k=0,1,...,K,d=1,2,...,q$, we assume
$\sup_s |X_d^{(k)}(s)|<\infty$.}
\end{assumption}

\begin{assumption}\label{assumpofsmooth}
\rm{Assume $\alpha_{d}^{(\mathcal{S},k)}\in \mathcal{C}^{\theta}(\mathcal{T})$, $\alpha_{d}^{(0)}\in \mathcal{C}^{\theta}(\mathcal{T})$ and $\alpha_{d}^{(\eta,k)}\in \mathcal{C}^{\theta_{\eta}}(\mathcal{T})$, $k=1,2...,K$.}
\end{assumption}

\begin{assumption}\label{assumpofmu}
\rm{
    (i) Each component of $\mathbb{E}\{ \bm{W}_i^{(k)}
          N [
\exp(\bm{W}_i^{(k)\top} \bm{\gamma}^{(k)}(\tau))]\},k=0,1,...,K$ is a Lipschitz function of $\tau$. 
(ii) Each component of $\mathbb{E}\{ \bm{W}_i^{(\eta)}
          N [
\exp(\langle \bm{X}_i^{(0)},\hat{\bm{\alpha}}^{(\mathcal{S})}\rangle+\bm{W}_i^{(\eta)\top} \bm{\gamma}(\tau))
     ]\}$ is a Lipschitz function of $\tau$.
}
\end{assumption}

\begin{assumption}\label{assumpofdensity}
\rm{
    $\tilde{f}(t|\bm{X})$ and $f(t|\bm{X})$ are uniformly bounded in $t$ and $\bm{X}$ for the target cohort and all the source cohorts.}
\end{assumption}

\begin{assumption}\label{assumofmatrix}
\rm{
For $k=0,1,2,...,K$, we assume
(i) $\mathbb{E}((\bm{W}_i^{(k)})^{\otimes 2})$ is positive define.
(ii) $\tilde{f}\{\exp(\bm{W}_i^{(k)}\bm{b})|
\\
\bm{X}_i^{(k)}\}>0$ for all $\bm{b}\in \mathcal{B}(\vartheta_{0,k},\sum_{d=1}^q M_{d,k}+qr)$. 
(iii) $\mathbb{E}\{ (\bm{W}_i^{(k)})^{\otimes 2}
          \bar{f} [
\exp(\bm{W}_i^{(k)}\bm{b})
     ]\times \exp(\bm{W}_i^{(k)}\bm{b})\}\times 
     \{\mathbb{E} \{
    (\bm{W}_i^{(k)})^{\otimes 2}
    \tilde{f} [
\exp(\bm{W}_i^{(k)}\bm{b})
]\times \exp(\bm{W}_i^{(k)}\bm{b})
     \}\}^{-1}$.
(iv) The minimal eigen-value of $\mathbb{E} \{
    (\bm{W}_i^{(k)})^{\otimes 2}
    \tilde{f} [
\exp\\
(\bm{W}_i^{(k)}\bm{\gamma}^{(k)})
]\times \exp\left(\bm{W}_i^{(k)}\bm{\gamma}^{(k)}\right)
     \}$ is bounded away from $0$.
}
\end{assumption}

\begin{assumption}\label{assumpofmatrixdiff}
\rm{
We assume
(i) $\mathbb{E}((\bm{W}_i^{(\eta)})^{\otimes 2})$ is positive define.
(ii) $\tilde{f}\{\exp(\langle \bm{X}_i^{(0)},\hat{\bm{\alpha}}^{(\mathcal{S})}\rangle+\bm{W}_i^{(\eta)}\bm{b})|
\bm{X}_i^{(0)}\}\\
>0$ for all $\bm{b}\in \mathcal{B}^{(\eta)}(\vartheta_{0,\eta},\sum_{d=1}^q M^{(\eta)}_{d}+qr)$. 
(iii) $\mathbb{E}\{ (\bm{W}_i^{(\eta)})^{\otimes 2}
          \bar{f} [
\exp(\langle \bm{X}_i^{(0)},\hat{\bm{\alpha}}^{(\mathcal{S})}\rangle+\bm{W}_i^{(\eta)}\bm{b})
     ]\times \exp(\langle \bm{X}_i^{(0)},
     \hat{\bm{\alpha}}^{(\mathcal{S})}\rangle+\bm{W}_i^{(\eta)}\bm{b})\}\times 
     \{\mathbb{E} \{
    (\bm{W}_i^{(\eta)})^{\otimes 2}
    \tilde{f} [
\exp(\langle \bm{X}_i^{(0)},\hat{\bm{\alpha}}^{(\mathcal{S})}\rangle+\bm{W}_i^{(\eta)}\bm{b})
]\times \exp(\langle \bm{X}_i^{(0)},
\hat{\bm{\alpha}}^{(\mathcal{S})}\rangle+\bm{W}_i^{(\eta)}\bm{b})
     \}\}^{-1}$
is uniformly bounded in $\bm{b}\in \mathcal{B}^{(\eta)}(\vartheta_{0,\eta},\sum_{d=1}^q M^{(\eta)}_{d}+qr)$.
(iv) The minimal eigen-value of $\mathbb{E} \{
    (\bm{W}_i^{(\eta)})^{\otimes 2}
    \tilde{f} [
\exp(\langle \bm{X}_i^{(0)},\hat{\bm{\alpha}}^{(\mathcal{S})}\rangle+\bm{W}_i^{(\eta)}\bm{\gamma}^{(\eta)})
]\times \exp(\langle \bm{X}_i^{(0)},\hat{\bm{\alpha}}^{(\mathcal{S})}\rangle+
\bm{W}_i^{(k)}\bm{\gamma}^{(\eta)})
     \}$ is bounded away from 0.
}
\end{assumption}

\begin{assumption}\label{assumpofkernel}
\rm{
    (i) The kernel function $\mathcal{K}(\cdot)$ is bounded and second-order differential continuous.
    (ii) Let $\|\bm{\mathcal{A}}_{\tau,L}\|= \max_j (\tau_j-\tau_{j-1})=o(1/\sqrt{\sup_{k=0,1,...,K}n_k})$.}
\end{assumption}

\begin{remark}
\rm{
Assumption \ref{assumpofmoment} places the restrictions on the moments of the functional predictors from the target cohort and all source cohorts.
Assumption \ref{assumpofsmooth} is about the smoothness of the functional coefficients from the target and all source cohorts.
Assumption \ref{assumpofmu} is the same as Assumption (A4) in \cite{jiang2020functional}.
Assumption \ref{assumpofdensity} imposes mild assumptions on the density functions related to the observed data from the target cohort and all source cohorts.
Assumption \ref{assumofmatrix} is the same as Assumptions (A5)-(A6) in \cite{jiang2020functional} and similar to the Conditions (C3)-(C4) in \cite{peng2008survival} ensuring the convergence rate obtained using each source cohort and target cohort. Assumption \ref{assumpofmatrixdiff} imposes assumption on the different parts between the target cohort and all source cohorts.
Assumption \ref{assumofmatrix} and Assumption \ref{assumpofmatrixdiff} imply that $\tilde{\bm{\mathcal{J}}}(\bm{\alpha}^{(0)}(\tau))$ is invertible.
For the kernel function used to calculate the weight of each source cohort, Assumption \ref{assumpofkernel}-(i) provides some regularity conditions. 
Assumption \ref{assumpofkernel}-(ii) specifies the order of the quantile grid size to derive the optimal nonparametric convergence of the estimators based on each source cohort and target cohort, which is similar to the condition of Theorem 2 in \cite{peng2008survival}. For example, we can let $\|\bm{\mathcal{A}}_{\tau,L}\|=O(1/\sqrt{\sup_{k=0,1,...,K}n_k})$.
}
\end{remark}

\subsection{Asymptotic Properties}
\label{asymptotic}
We state the following theorems, whose detailed proofs are provided in the supplementary document. Theorem \ref{weighttheory} demonstrates the convergence rate of the calculated similarity weight in \texttt{SITL} Algorithm \ref{altrans3}.

\begin{theory} 
\label{weighttheory}
\rm{
        Under Assumptions \ref{assumpofmoment}, \ref{assumpofsmooth}, \ref{assumpofmu}-(i), \ref{assumpofdensity}, \ref{assumofmatrix} and \ref{assumpofkernel}, 
        if $M_{d,k} = O(n_k^{1/(2\theta+1)})$
        and $M_{d} = O(n_0^{1/(2\theta+1)})$ as $n_k,n_0\to \infty$, for $k=1,2,...,K$, we obtain
\begin{equation}\label{weight}
   \begin{split}
       & |\hat{\omega}^{(k)}-\omega^{(k)}|\\
       = &O_p\left(
        \sqrt{\frac{\|\bm{\mathcal{A}}_{\tau,L}\|}{h^2}}\left( n_{k}^{-\frac{\theta}{2\theta+1}} +
         n_{0}^{-\frac{\theta}{2\theta+1}} 
        +\sqrt{\frac{1}{n_0}}
        \sup_{d,\tau}\|\alpha^{(0)}_d(\cdot,\tau)-\alpha_{d}^{(\mathcal{S},k)}(\cdot,\tau)\|_{L_2}\right)
        \right).\\
    \end{split}
\end{equation}
Usually we have $n_k >>n_0$, then
\begin{align*}
    & |\hat{\omega}^{(k)}-\omega^{(k)}|=O_p\left(
        \sqrt{\frac{\|\bm{\mathcal{A}}_{\tau,L}\|}{h^2}}\left( n_{0}^{-\frac{\theta}{2\theta+1}} + \sqrt{\frac{1}{n_0}}
        \sup_{d,\tau}\|\alpha^{(0)}_d(\cdot,\tau)-\alpha_{d}^{(\mathcal{S},k)}(\cdot,\tau)\|_{L_2}\right)
        \right).
\end{align*}
Further, when $\sup_{d,\tau}\|\alpha^{(0)}_d(\cdot,\tau)-\alpha_{d}^{(\mathcal{S},k)}(\cdot,\tau)\|_{L_2} = o(n_0^{1/(4\theta+2)})$, 
$$|\hat{\omega}^{(k)}-\omega^{(k)}|=O_p\left(\sqrt{\frac{\|\bm{\mathcal{A}}_{\tau,L}\|}{h^2}}n_{0}^{-\frac{\theta}{2\theta+1}}\right).$$
    }
\end{theory}

\begin{remark}
\rm{
Theorem \ref{weighttheory} indicates that the errors in the estimated weights decrease as $\|\bm{\mathcal{A}}_{\tau,L}\|$ decreases. Additionally, the error diminishes as $n_0$ increases and as the difference between the target and source cohorts reduces.
Regarding the bandwidth $h$, the error decreases as $h$ increases.
}
\end{remark}

The following Theorem \ref{transfertheory}
establishes the convergence rate of the functional coefficient estimators obtained through the \texttt{SITL} method.
\begin{theory}
\label{transfertheory}
\rm{
Under Assumptions \ref{assumpofmoment}-\ref{assumpofkernel}, if $M_{d,k} = O(n_k^{1/(2\theta+1)})$, $M_{d} = O(n_0^{1/(2\theta+1)})$,
$M_{d}^{(\eta)} = O(n_0^{1/(2\theta_{\eta}+1)})$,
and $n_k>>n_0, k=1,2...,K$ as $n_k,n_0\to \infty$, we have
\begin{equation}\label{transupper}
\begin{split}
     & \sup_{\substack{\tau\in[0,\tau_{U}]
    }}
     \|\tilde{\alpha}_d(\cdot,\tau)-\alpha_{d}^{(0)}(\cdot,\tau)\|_{L_2} \\ = & 
     O_p\left(n_{0}^{-\frac{\theta_{\eta}}{2\theta_{\eta}+1}}+\sup_{k}  n_{k}^{-\frac{\theta}{2\theta+1}} + \sqrt{\|\bm{\mathcal{A}}_{\tau,L}\|}\left( 
         n_{0}^{-\frac{\theta}{2\theta+1}} 
        +\sqrt{\frac{1}{n_0}}
        \sup_{k,d,\tau}\|\alpha^{(0)}_d(\cdot,\tau)-\alpha_{d}^{(\mathcal{S},k)}(\cdot,\tau)\|_{L_2}\right)\right).\\
\end{split} 
\end{equation}
}
\end{theory}

\begin{remark}
\rm{
Moreover, we demonstrate the effectiveness of the proposed \texttt{SITL} for mFCQR by analyzing the convergence rate and comparing it with that of classical estimators based solely on the target cohort. The convergence rate of the estimators based on the target cohort is $O_p(n_{0}^{-\theta/(2\theta+1)})$.
For simplicity, we let $\sup_{k,d,\tau}\|\alpha^{(0)}_d(\cdot,\tau)-\alpha_{d}^{(\mathcal{S},k)}(\cdot,\tau)\|_{L_2} = o(n_0^{1/(4\theta+2)})$
and $\|\bm{\mathcal{A}}_{\tau,L}\|=O(1/\sup_k n_k)$, then the convergence rate (\ref{transupper}) of the estimator by our proposed transfer learning method is given by $$O_p\left(n_{0}^{-\frac{\theta_{\eta}}{2\theta_{\eta}+1}}+\sup_{k}  n_{k}^{-\frac{\theta}{2\theta+1}} + \sqrt{\frac{1}{\sup_k n_k}} 
         n_{0}^{-\frac{\theta}{2\theta+1}}
        \right)= O_p\left(n_{0}^{-\frac{\theta_{\eta}}{2\theta_{\eta}+1}}+\sup_{k}  n_{k}^{-\frac{\theta}{2\theta+1}}
        \right),$$
In the scenario where $\theta{\eta}\leq \theta$, the use of source cohorts and transfer learning becomes ineffective.
Essentially, when $\theta{\eta}\leq \theta$, our method cannot outperform the classical estimators. The effectiveness of the proposed \texttt{SITL} is therefore justifiable only when $\theta{\eta}> \theta$. 

In the second scenario, when $\theta_{\eta}$ satisfies $n_{0}^{-\theta_{\eta}/(2\theta_{\eta}+1)}>\sup_{k} n_{k}^{-\theta/(2\theta+1)}$, 
the estimators obtained from the proposed transfer learning method achieve a convergence rate of $n_{0}^{-\theta_{\eta}/(2\theta_{\eta}+1)}>>n_{0}^{-\theta/(2\theta+1)}$.
If $\theta_{\eta}$ is large enough, for example, when $\theta_{\eta}=\infty$
(indicating the difference function between all source cohorts and the target cohort are sufficiently simple and smooth), 
the convergence rate of our method is given by
$n_0^{-1/2}$ under $n_0<\sup_k n_k^{2\theta/(2\theta+1)}$, which is the optimal convergence rate of parametric estimators.
}
\end{remark}

We next establish the asymptotic normality of the estimators in Theorem \ref{theorynormal}.

\begin{theory}
\label{theorynormal}
\rm{
Without loss of generality, and to simplify the expression, we let $M^{(\eta)}_d \equiv M^{(\eta)}$.
Under Assumptions \ref{assumpofmoment}-\ref{assumpofkernel}, if $n_k>>n_0, k=1,2...,K$, $M_{d,k} = O(n_k^{1/(2\theta+1)})$, $M_{d} = O(n_0^{1/(2\theta+1)})$, $\sqrt{\|\bm{\mathcal{A}}_{\tau,L}\|}
         n_{0}^{-\theta/(2\theta+1)}= o(\sqrt{M^{(\eta)}/n_0})$,
$\sqrt{\|\bm{\mathcal{A}}_{\tau,L}\|/{n_0}}
        \sup_{k,d,\tau}\|\alpha^{(0)}_d(\cdot,\tau)-\alpha_{d}^{(\mathcal{S},k)}(\cdot,\tau)\|_{L_2}= o(\sqrt{M^{(\eta)}/n_0})$,
$\sup_{k} n_{k}^{-\theta/(2\theta+1)} = o(\sqrt{M^{(\eta)}/n_0})$ and 
$(M^{(\eta)})^{(2\theta_{\eta}+1)}/n_0\to \infty$ as $n_k,n_0\to \infty$, we have
\begin{align}\label{theoryalpha}
   & \sqrt{\frac{n_0}{M^{(\eta)}}} \left\{\tilde{\alpha}_d(s,\tau) - \alpha_{d}^{(0)}(s,\tau) \right\} 
   \nonumber
   \\
    \to & 
    (M^{(\eta)})^{-1}
    \bm{B}^{(\eta)\top}_{d}(s)
    \bm{A}_d \left[\bm{\mathcal{J}}\left(
    \bm{\alpha}^{(0)}(\tau)
    \right)\right]^{-1}\phi \left[\bm{G}(\tau), \tilde{\bm{\mathcal{J}}}\left(
    \bm{\alpha}^{(0)}(\tau)
\right),\left[\bm{\mathcal{J}}\left(
    \bm{\alpha}^{(0)}(\tau)
    \right)\right]^{-1} \right],
\end{align}
where $\bm{A}_d = (\bm{0}_{M^{(\eta)}\times (M^{(\eta)}(d-1))},\bm{I}_{M^{(\eta)}},\bm{0}_{M^{(\eta)}\times (M^{(\eta)}(q-d))})$, $\bm{0}_{M^{(\eta)}\times (M^{(\eta)}(d-1))}$ is a $M^{(\eta)}\times (M^{(\eta)}(d-1))$ dimensional matrix of zeros, $\bm{I}_{M^{(\eta)}}$ denotes a $M^{(\eta)}\times M^{(\eta)}$ identity matrix and $\bm{G}(\tau)$ is a tight zero-mean Gaussian process with covariance $\bm{\Sigma}(\tau_l,\tau_{l'}) = M^{(\eta)}\mathbb{E}(\bm{\iota}_i(\tau_l),\bm{\iota}_i(\tau_{l'}))=O(1)$.
}
\end{theory}

\begin{remark}
\rm{
    In Theorem \ref{theorynormal}, the assumption regarding the number of knots used in the Debias Step ensures that the bias introduced by the Transfer Step and B-spline approximation becomes asymptotically negligible. This simply means that the number of knots required is larger than what is necessary to achieve the optimal convergence rate, as demonstrated in Theorem \ref{transfertheory}.
    Specifically, the condition $(M^{(\eta)})^{(2\theta_{\eta}+1)}/n_0\to \infty$ indicates that the number of knots required to make the B-spline approximation error asymptotically negligible in our method is still smaller than the condition $M^{(2\theta+1)}/n_0\to \infty$ required by estimators based solely on the target cohort.
}
\end{remark}

\subsection{Resampling \texttt{SITL} Methods}
\label{CI}
To make inferences about $\alpha_{d}^{(0)}(\cdot,\cdot)$, we need to know the true covariance function of $\tilde{\alpha}_{d}^{(0)}(\cdot,\cdot)$. However, as indicated by Theorem \ref{theorynormal}, the covariance function involves integrals, product integral,
the unknown density functions $f(t|\bm{X^{(0)}})$ and $\tilde{f}(t|\bm{X^{(0)}})$, making it challenging to calculate and estimate. \cite{peng2008survival} proposed using a resampling approach to estimate the covariance matrix based solely on the target cohort. However, when the sample size of the target cohort is small, the estimators obtained from each resampling run tend to be unstable, resulting in a not good enough covariance estimator.

Intuitively, we aim to use the \texttt{SITL} approach to estimate the covariance, which effectively combines the advantages of both our proposed transfer learning and the resampling method. The details of the point-wise confidence interval construction procedure are provided in Algorithm \ref{altrans4}. In this algorithm, the transfer step follows the same procedure as in Algorithm \ref{altrans3}. In the Debias Step, we incorporate the resampling approach by introducing a stochastic perturbation to the Debias Step from Algorithm \ref{altrans3}. Specifically, we generate independent variates $\zeta_1,...,\zeta_{n_0}$ from a nonnegative known distribution with mean 1 and variance 1. For each quantile level $\tau_j$, the estimation equation in the Debias Step is updated to (\ref{bootdebais}).

In Theorem \ref{theoryboot}, we present the theoretical foundation for the confidence interval construction procedure outlined in Algorithm \ref{altrans4}. This theorem guarantees that the $(1-a\%)$ confidence interval for each functional coefficient in Algorithm \ref{altrans4} is approximately at the $(1-a\%)$ level. 

\begin{theory}
\label{theoryboot}
\rm{
Under the assumptions in Theorem \ref{theorynormal},
the distribution of $\sqrt{n_0/{M^{(\eta)}}} \{\tilde{\alpha}^*_d(s,\tau)- \tilde{\alpha}_{d}(s,\tau) \}$ is asymptotically equivalent to the unconditional distribution of 
$\sqrt{n_0/{M^{(\eta)}}} \{\tilde{\alpha}_d(s,\tau) - \alpha_{d}^{(0)}(s,\tau)\}$ as shown in Theorem \ref{theorynormal}. Thus, the variance of 
$\tilde{\alpha}_{d}(s,\tau)$ 
can be estimated by the sample variance of $\tilde{\bm{\alpha}}^{*(b)}(s,\tau)\}_{b=1}^B$ for a fixed $s\in \mathcal{T}$ and $\tau\in(0,\tau_U]$.
}
\end{theory}

\begin{algorithm}[htbp]
\footnotesize
\SetKwData{Left}{left}\SetKwData{This}{this}\SetKwData{Up}{up}
\SetKwFunction{Union}{Union}\SetKwFunction{FindCompress}{FindCompress}
\SetKwInOut{Input}{Input}\SetKwInOut{Output}{Output}
\Input{Target data, all source data, tunning parameters and quantile levels}

\BlankLine
\begin{itemize}

       		  \item \textbf{Transfer Step}: 
                    Take a kernel-based weighted average of $\hat{\bm{\alpha}}^{(\mathcal{S},k)}(s,\tau)$ and denote by $\hat{\bm{\alpha}}^{(\mathcal{S})}(s,\tau)$:
                    \begin{align*}
                        \hat{\bm{\alpha}}^{(\mathcal{S})}(s,\tau) & =  \frac{1}{\sum_{k=1}^K n_k} \sum_{k=1}^K n_k \hat{\omega}_k \hat{\bm{\alpha}}^{(\mathcal{S},k)}(s,\tau),
                    \end{align*}
                    where the weight $\hat{\omega}_k$ is calculated as introduced in the first step of Algorithm \ref{altrans3}.
			
			\item \textbf{Debias Step:} 
Let $\zeta_1,\cdots, \zeta_n$ be independent variates from a nonnegative
known distribution with mean 1 and variance 1, for example,
exponential (1).
   Estimate the differences ${\bm{\alpha}}^{*(\eta)}(s,\tau)$ 
    by solving the following equation with respect to $\bm{\gamma}^{(\eta)}(\tau_j)$ at given quantile level $\tau_j$:
\begin{align}\label{bootdebais}
       & \sum_{i=1}^n \zeta_i \bm{W}_i^{(\eta)}
        \Bigg\{ I \left[
    \hat{e}_i^{(\mathcal{S})}(\tau_j)-\bm{W}_i^{(\eta)\top} \bm{\gamma}^{(\eta)}(\tau_j)\leq 0; \delta_i=1
     \right]-u_{ij}^*
     \Bigg\},
\end{align}
with $u_{ij}^*=\sum_{l=0}^{j-1} I\{\hat{e}_i^{(\mathcal{S})}(\tau_l)\geq \bm{W}_i^{(\eta)\top} \hat{\bm{\gamma}}^{(*\eta)}(\tau_l) \} \times  \{H(\tau_{l+1})-H(\tau_l)\}$.\\
Then we can get the 
$\tilde{\bm{\alpha}}^*(s,\tau)= \hat{\bm{\alpha}}^{(\mathcal{S})}(s,\tau)+\hat{\bm{\alpha}}^{(*\eta)}(s,\tau)$.\\
We repeat the \textbf{Debias Step} $B$ times 
and obtain $\tilde{\bm{\alpha}}^{*(b)}(s,\tau)\}_{b=1}^B$.
\end{itemize}
\BlankLine
\Output{
The CI for $\alpha_d(s,\tau),d=1,2...,q$ at fixed quantile level $\tau$ and fixed point $s$ is: 
$$\left[
\tilde{\alpha}_d(s,\tau)-q_{a/2}\times \mathrm{SD}(
\tilde{\alpha}_d^{*(b)}(s,\tau)), \tilde{\alpha}_d(s,\tau)+q_{a/2}\times \mathrm{SD}(
\tilde{\alpha}_d^{*(b)}(s,\tau))
\right],$$
where $q_{a/2}$ is the $a/2$-quantile of the standard normal distribution.
}
\caption{
Confidence Interval Constructions Via the Resampling \texttt{SITL} Algorithm for mFCQR
}\label{altrans4}
\end{algorithm}

\section{Simulation Studies}
\label{simu}
In this section, we present simulation studies to assess the finite sample performance of our proposed Similarity-Informed Transfer Learning (\texttt{SITL}) method. Additionally, we compare its performance with three alternative conventional methods: (i) using only the target cohort data, (ii) pooling target and source cohort data, and (iii) a hard-threshold transfer learning approach.

\subsection{Simulation Configurations}
For all cohorts, we consider two functional predictors $\bm{X}_i(s)=(X_{i1}(s),X_{i2}(s))^{\top}$, where $X_{1i}(s)$ is constructed by $X_{1i}(s)=\left|\sum_{k=1}^K\xi_k U_{ik}\phi_k(s)\right|, \phi_1(s) =1, \phi_k(s)=\sqrt{2}\cos\left\{(k-1)\pi s\right\}$ for $k>1$, with $\xi_k=(-1)^{k+1}k^{-1}$, $U_{ik}\overset{iid}{\sim} U[-\sqrt{3},\sqrt{3}]$, $K=20$, and $X_{2i}(s)$ is given by $X_{2i}(s) = \sum_{k=1}^K\zeta_{ik} B_k(s)$, $\zeta_{ik}\overset{iid}{\sim} N(0,\sigma_{\zeta}^2)$, and $B_k(t)$ is the cubic B-spline basis function defined on $K-4$ equally spaced interior knots in $[0,1]$. The data of the target cohort is generated as 
\begin{equation*}
    \log(T_i) = \int_{0}^1X_{1i}(s)\psi_{11}(s,\tau)\mathrm{d}s+\int_{0}^1X_{2i}(s)\psi_{2}(s,\tau)\mathrm{d}s
    +\left\{\int_{0}^1X_{1i}(s)\psi_{12}(s)\mathrm{d}s\right\}\epsilon_i,
    \end{equation*}
then $\alpha_1(s,\tau) = \psi_{11}(s,\tau)+\psi_{12}(s)F_{\epsilon}^{-1}(\tau)$,
and $\alpha_2(s,\tau) = \psi_2(s,\tau)$, where $\psi_{11}(s,\tau) = 4\sum_{k=1}^K(-1)^k\\
k^{-2}\left\{\sqrt{2}\cos \left[(k-1)\pi s\right]\right\}$, $\psi_{12}(s)=1$,  $\alpha_{2}(s,\tau) = 4 \left\{\cos(3\pi s)\right.\left.+\sin(3\pi s)\right\}$,  $\epsilon_i\overset{iid}{\sim}N(0,\sigma_{\epsilon}^2)$, and $\sigma_{\epsilon}^2=0.2$. 
For the source cohorts, we consider the following four scenarios:
\begin{itemize}
        \item \textbf{Case 1}: The source cohorts use the same
        functional coefficients $\alpha_{1}^{(\mathcal{S})}(s,\tau)$, $\alpha_{2}^{(\mathcal{S})}(s,\tau)$ as those of the target cohort.
        \item \textbf{Case 2}: The source cohorts use the same first functional coefficient $\alpha_{1}^{(\mathcal{S})}(s,\tau)$ as that of the target cohort and a different second functional coefficient $\alpha_{2}^{(\mathcal{S})}(s,\tau)=4\{\cos(3\pi s)+\sin(3\pi s)\}+ 10\exp(s)$. Note $\alpha^{(\mathcal{S})}_{2}(s,\tau) = \alpha_{2}(s,\tau)+10\exp(s)$.
\item \textbf{Case 3}:
        The source cohorts use the same first functional coefficient $\alpha_{1}^{(\mathcal{S})}(s,\tau)$ as that of the target cohort and a different second functional coefficient $\alpha_{2}^{(\mathcal{S})}(s,\tau)=12 \{\cos(3\pi s)+\sin(3\pi s)\}$.
        Note $\alpha^{(\mathcal{S})}_{2}(s,\tau) = 3\alpha_{2}(s,\tau)$.
    \item \textbf{Case 4}: The source cohorts use a different first functional coefficient  $\alpha_{1}^{(\mathcal{S})}(s,\tau)=3\psi_{11}^{(\mathcal{S})}(s,\tau)\\+\psi_{12}(s)F_{\epsilon}^{-1}(\tau)$, where $\psi_{11}^{(\mathcal{S})}(s,\tau) = 12\sum_{k=1}^K(-1)^kk^{-2}\left\{\sqrt{2}\cos \left[(k-1)\pi s\right]\right\}$, and a different second functional coefficient  $\alpha_{2}^{(\mathcal{S})}(s,\tau)=12 \left\{\cos(3\pi s)+\sin(3\pi s)\right\}+ 10\exp(s)$. Note $\alpha^{(\mathcal{S})}_{2}(s,\tau) = 3\alpha_{2}(s,\tau)+10\exp(s)$.
    \end{itemize}

These four cases represent four different scenarios regarding the disparity between the source cohorts and the target cohort. The first case represents independent and identically distributed data (i.i.d.) with the target cohort, making it the most relevant and valid source of information.
Regardless of the methods, information from the source cohorts should be utilized to its fullest potential. In the second case, the functional coefficient $\alpha_2^{\mathcal{(S)}}(s,\tau)$ differs from  $\alpha_2(s,\tau)$ of the target cohort by $10\exp(s)$, indicating a considerable disparity with the target cohorts, but still retaining some useful information. Thus, it is reasonable to assign a small but non-zero weight to the source cohorts. Addressing how to identify and effectively utilize information from source cohorts like the second case is a key challenge of our study. In contrast, the third case's $\alpha_2^{\mathcal{(S)}}(s,\tau)$ differs only in its coefficient from the target cohort, representing minimal data disparity and serving as a valid source of information. Finally, the fourth case involves substantial difference in both $\alpha_1^{\mathcal{(S)}}(s,\tau)$ and $\alpha_2^{\mathcal{(S)}}(s,\tau)$ compared to the target cohort. 
Over-reliance on information from such source cohorts would negatively affect the estimation accuracy and should therefore be avoided.

We perform simulation studies by varying the sample size of the source cohorts to $\bm{n}_1=(500,1000,\\500,1000)$, the number of source cohorts to $K=(4,8,12,16)$, and the censoring rate of the source cohorts to $(0\%,10\%,20\%,30\%)$). The sample size for the target cohort is $n_0=(100,150,200,250)$. The bandwidth for the Gaussian kernel function used to define the similarity weights in (\ref{diff}) is set to $h=2\log(5n_0)$. All simulation results presented are based on 100 simulation replications.

\subsection{Simulation Results}
To evaluate the performance of the estimators, we use the root mean squared errors (RMSEs) for each functional coefficient, which is defined as 
\begin{align*}\text{RMSE}(\hat{\alpha}_d(\tau))=\sqrt{\int_0^1\left[\hat{\alpha}_d(s,\tau)-\alpha_d(s,\tau)\right]^2ds}.
\end{align*} 
Table \ref{ta:censor rate} summarizes the simulation results of the two estimated functional coefficients $\hat{\alpha}_d(s,\tau)$, $d=1,2,$ under various censoring rates when the quantile $\tau=0.5$, $n_0=100$ and the number of source cohorts $K=4$. 
Note that the results using our \texttt{SITL} method outperform the other three methods for all censoring rates. More specifically, in the estimation of $\alpha_1$, compared to the method only using the target cohort, our \texttt{SITL} method improves \text{RMSE} by at least 25\%. For the estimation of $\alpha_2$, our \texttt{SITL} method achieves approximately a 10\% improvement over the method only using the target cohort, and a even more significant enhancement when compared to the method of pooling target and source cohort data and the hard-threshold transfer learning method, with improvements ranging from 60\% to 95\%. This is because, in our setup, the difference in $\alpha_2$ between the source and target cohorts is more pronounced, while the difference for $\alpha_1$ is relatively smaller. This indicates that the greater the difference between the source and target cohorts, the more evident the superiority of our proposed \texttt{SITL} method becomes. Compared to other methods, our approach provides more robust estimates and is better suited for analysis in practical application scenarios. 
\begin{table}[htbp]
\centering
  \fontsize{10}{12}\selectfont
  \begin{threeparttable}
  \caption{The mean and standard errors (in parenthesis) of the root mean squared errors (RMSE) for the estimated  coefficient function $\alpha_d(s,\tau),d=1,2$, respectively,
   under the quantile $\tau=0.5$, $n_0=100$, $k=4$ and varying censoring rates using our proposed Similarity-Informed Transfer Learning (\texttt{SITL}) method and three alternative conventional methods including (i) using the target cohort data only (called \texttt{Target}), (ii) pooling target and source cohorts together (called \texttt{Pooled}), and (iii) the hard-threshold transfer learning method (called \texttt{Transfer\_HT}).}
  \label{ta:censor rate}
\renewcommand\arraystretch{0.8}
  \setlength{\tabcolsep}{3mm}{
\begin{tabular}{cccccc}
\toprule

\multirow{2}{*}{}&\multirow{2}{*}{\textbf{Methods}}&\multicolumn{4}{c}{\textbf{Censoring Rate}}\\
\cline{3-6}
 
  &  & \textbf{0\%} & \textbf{10\%} & \textbf{20\%} &
  \textbf{30\%}\\
  \midrule
\multirow{4}{*}{\textbf{RMSE}($\hat{\alpha}_1$)}& 
 \textbf{Target}
 & 0.742(0.268) 
 & 0.787 (0.269) 
 & 0.891 (0.323)
 & 0.894 (0.324) 
   \\ 
 & \textbf{Pooled}
 & 2.839(0.492) 
 & 2.822 (0.543)
 & 2.157 (0.463)
 & 1.969 (0.429) 
    \\ 
  & \textbf{Transfer\_HT}
  & 0.577(0.244) 
  & 0.674 (0.254)
  & 0.710 (0.282)
  & 0.699 (0.240) 
     \\ 
  & \textbf{SITL}
  & 0.518(0.212) 
  & 0.535 (0.194)
  & 0.568 (0.208)
  & 0.666 (0.230) 
    \\

    \midrule
\multirow{4}{*}{\textbf{RMSE}($\hat{\alpha}_2$)}& \textbf{Target}
  & 0.710(0.074) 
  & 0.718 (0.095) 
  & 0.771 (0.118)
  & 0.789 (0.134) 
    \\ 
  & \textbf{Pooled}
  & 11.223(0.486) 
  & 11.662 (0.797)
  & 13.713 (0.776)
  & 14.041 (0.620) 
     \\ 
  & \textbf{Transfer\_HT} 
  & 1.859(0.315) 
  & 2.778 (0.527) 
  & 2.506 (0.440)
  & 2.164 (0.406) 
     \\ 
  & \textbf{SITL}
  & 0.662 (0.055) 
  & 0.669 (0.051) 
  & 0.689 (0.065)
  & 0.709 (0.086) 
     \\
  
   \bottomrule
\end{tabular}}
\end{threeparttable}
\end{table}

Table \ref{ta:loss} shows that our \texttt{SITL} method also outperforms other methods in terms of prediction errors, achieving superior predictive performance. Here, we calculate prediction error using the loss function (\ref{loss_tar}), with the subset $\mathcal{I}^c$ replaced by the test set. 
Our \texttt{SITL} method achieves a significant reduction in prediction error, outperforming other approaches across various scenarios. When four source cohorts are available, it reduces prediction error by 89.0\% compared to the method that pools target and source cohort data. Additionally, it achieves a 64.2\% lower prediction error than the hard-threshold transfer learning method under the same conditions. Compared to using only the target cohort, the \texttt{SITL} method delivers a 4.7\% reduction in prediction error. These improvements remain consistent when the number of source cohorts increases to 16.

\begin{table}[htbp]
\centering
  \fontsize{10}{12}\selectfont
  \begin{threeparttable}
  \caption{The mean and standard errors (in parenthesis) of the prediction errors
   when the quantile $\tau=0.5$, $n_0=100$, and the censoring rate $c=10\%$ and varying number of source cohorts using our proposed Similarity-Informed Transfer Learning (\texttt{SITL}) method and three alternative conventional methods including (i) using the target cohort data only (called \texttt{Target}), (ii) pooling target and source cohorts together (called \texttt{Pooled}), and (iii) the hard-threshold transfer learning method (called \texttt{Transfer\_HT}).}
  \label{ta:loss}
\renewcommand\arraystretch{0.8}
  \setlength{\tabcolsep}{5mm}{
\begin{tabular}{ccccc}
\toprule

\multirow{1}{*}{\textbf{Number of}}&\multicolumn{4}{c}{\textbf{Methods}}\\
\cline{2-5}
 
 \textbf{Source Cohorts}& \textbf{Target} & \textbf{Pooled} & \textbf{Transfer\_HT} &
  \textbf{\texttt{SITL}}\\
  \midrule
 
 \textbf{$K=4$} 
 & 0.256 (0.028) 
& 2.227 (0.220)
& 0.682 (0.072)
& 0.244 (0.026)
    \\ 
  \textbf{$K=16$} 
 & 0.248 (0.031)
 & 2.036 (0.200)
 & 0.601 (0.087)
 & 0.238 (0.026)
\\
  
   \bottomrule
\end{tabular}}
\end{threeparttable}
\end{table}

Table \ref{ta:number} presents the \text{RMSEs} of the estimated functional coefficients obtained using the \texttt{SITL} method with $\tau = 0.5$ and a censoring rate of $c = 10\%$, across varying sample sizes of the target cohorts and numbers of source cohorts. The \text{RMSEs} for both functional coefficients decrease significantly as the target cohort sample size increases and as the number of source cohorts $K$ grows. This aligns with the theoretical results discussed in Section 3. Notably, even when the target cohort sample size is as small as $100$ and the number of source datasets is limited to $4$, the \texttt{SITL} method demonstrates satisfactory estimation performance. This highlights the robustness of our method, particularly in practical applications where real-world datasets often have limited sample sizes.

\begin{table}[htbp]
\centering
  \fontsize{10}{12}\selectfont
  \begin{threeparttable}
  \caption{The mean and standard errors (in parenthesis) of the root mean squared errors (RMSE) for the estimated coefficient function $\alpha_d(s,\tau),d=1,2$, respectively,
   with the quantile $\tau=0.5$ and the censoring rate $c=10\%$, when varying the sample size of the target cohort and the number of source cohorts using the proposed Similarity-Informed Transfer Learning (\texttt{SITL}) method.}
  \label{ta:number}
\renewcommand\arraystretch{0.8}
  \setlength{\tabcolsep}{3mm}{
\begin{tabular}{ccccc}
\toprule
\multirow{1}{*}{}&\multicolumn{4}{c}{\textbf{Different Sample size of Target Cohort}}\\
\cline{2-5}
 
  & \textbf{$(n_0,K)=(100,4)$} & \textbf{$(n_0,K)=(150,4)$} & \textbf{$(n_0,K)=(200,4)$} & \textbf{$(n_0,K)=(250,4)$}\\
  \midrule
  {\textbf{RMSE}($\hat{\alpha}_1$)}
 & 0.540 (0.195) 
 & 0.469 (0.157) 
 & 0.429 (0.145)
 & 0.416 (0.140)
  \\
{\textbf{RMSE}($\hat{\alpha}_2$)}
 & 0.678 (0.070) 
 & 0.664 (0.056) 
 & 0.644 (0.041) 
 & 0.636 (0.040)
 \\ 
 \hline
\multirow{1}{*}{}&\multicolumn{4}{c}{\textbf{Different Number of Source Cohorts}}\\
\cline{2-5}
 
  & \textbf{$(n_0,K)=(100,4)$} & \textbf{$(n_0,K)=(100,8)$} & \textbf{$(n_0,K)=(100,12)$} & \textbf{$(n_0,K)=(100,16)$}\\
  \midrule
{\textbf{RMSE}($\hat{\alpha}_1$)}
 & 0.537 (0.197) 
 & 0.504 (0.175) 
 & 0.471 (0.178)
 & 0.433 (0.175)
  \\
{\textbf{RMSE}($\hat{\alpha}_2$)}
 & 0.680 (0.065) 
 & 0.676 (0.058) 
 & 0.673 (0.053)
 & 0.662 (0.051)
 \\ 
  
   \bottomrule
\end{tabular}}
\end{threeparttable}
\end{table}

\begin{figure}[h]
	\centering
	\subfigure[Target, $\hat{\alpha}_1(s,\tau)$]{\includegraphics[width=4.5cm]{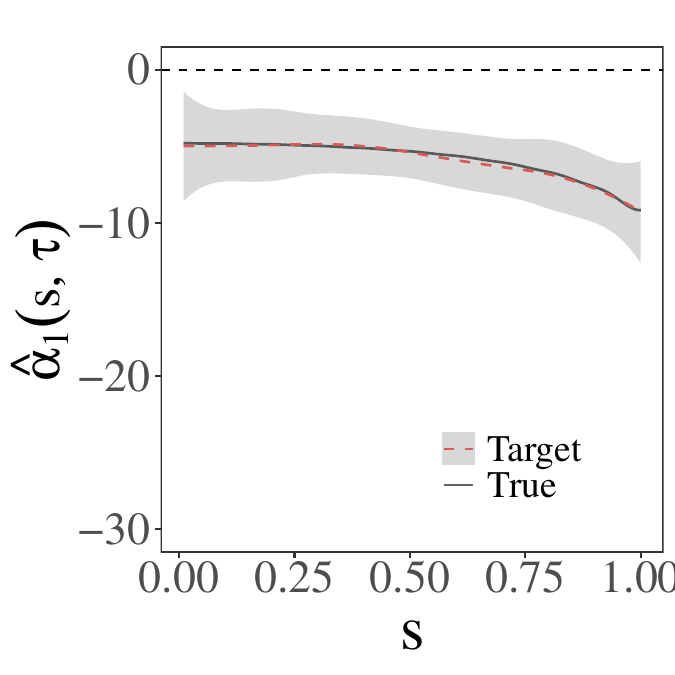}}
	\subfigure[Pooled, $\hat{\alpha}_1(s,\tau)$]{\includegraphics[width=4.5cm]{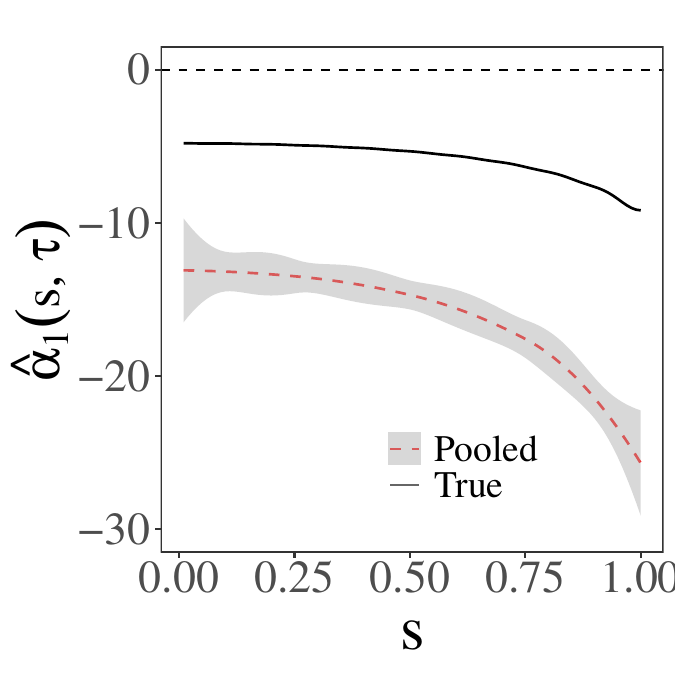}}
    \subfigure[\texttt{SITL}, $\hat{\alpha}_1(s,\tau)$]{\includegraphics[width=4.5cm]{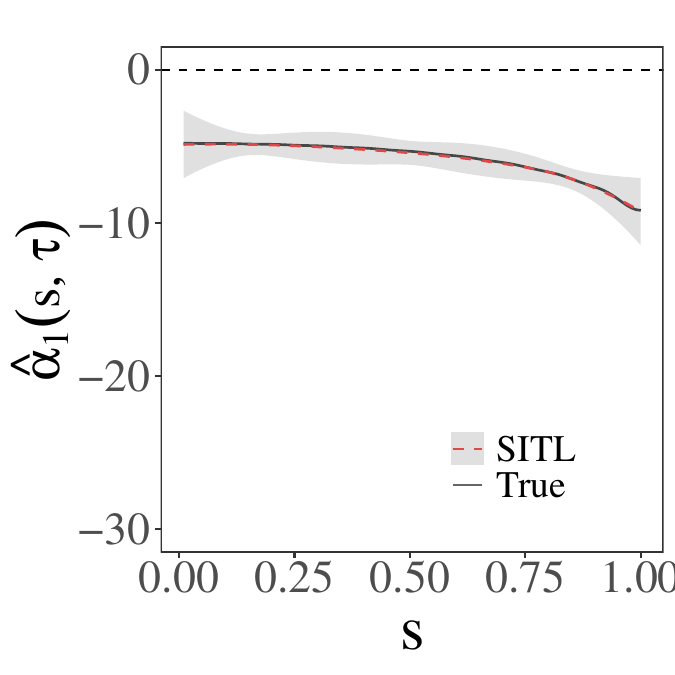}}
	\\
    \subfigure[Target, $\hat{\alpha}_2(s,\tau)$]{\includegraphics[width=4.5cm]{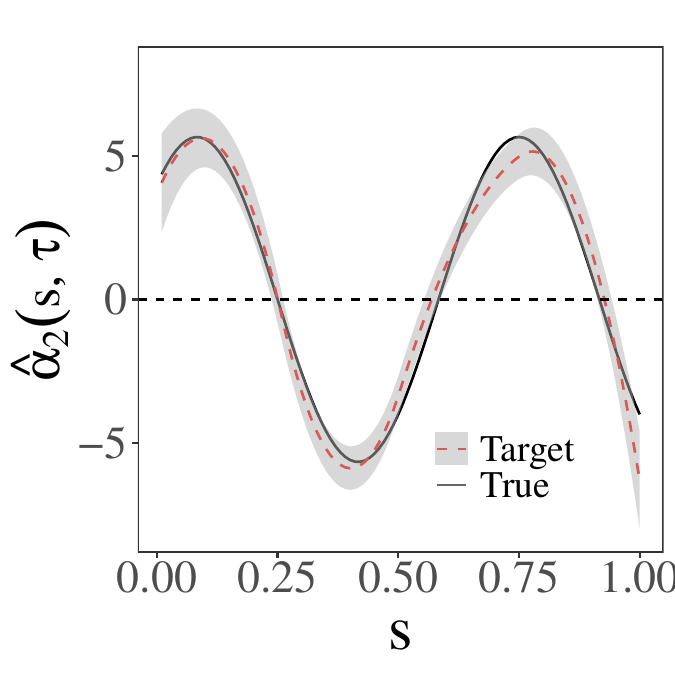}}
    \subfigure[Pooled, $\hat{\alpha}_2(s,\tau)$]{\includegraphics[width=4.5cm]{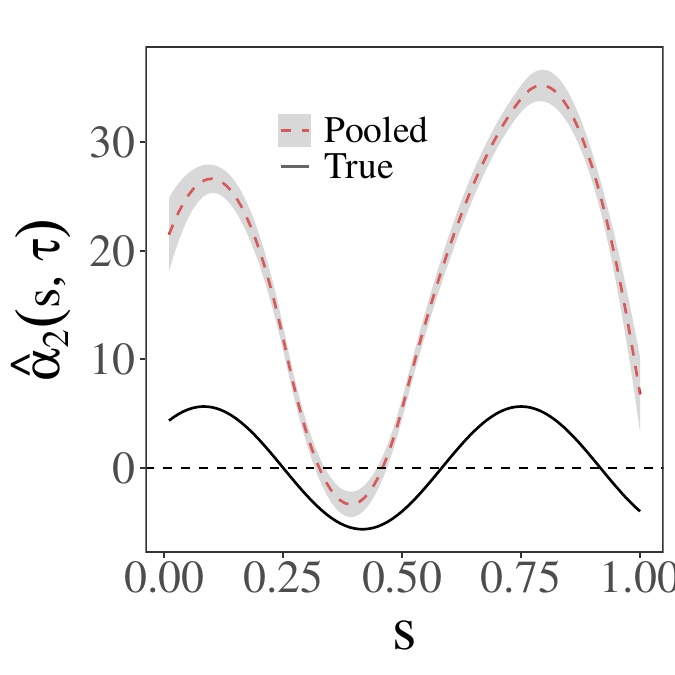}}
	\subfigure[\texttt{SITL}, $\hat{\alpha}_2(s,\tau)$]{\includegraphics[width=4.5cm]{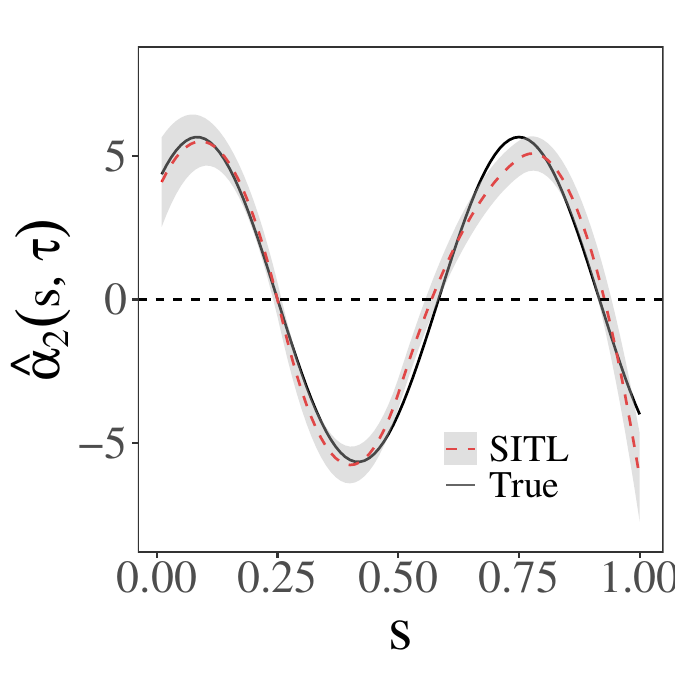}}
	\caption{The true functional coefficients ${\alpha}_d(s,\tau)$, $d=1,2$, the estimated functional coefficients and the corresponding 95\% point-wise confidence intervals when the quantile $\tau=0.3$, the censoring rate $c=10\%$, the sample size of the target cohort $n_0=100$, and the number of source cohorts $K=4$ using our proposed Similarity-Informed Transfer Learning (\texttt{SITL}) method and two alternative conventional methods including (i) using the target cohort data only (called \texttt{Target}) and (ii) pooling target and source cohorts together (called \texttt{Pooled}). }\label{sim:fig1}
\end{figure}

We now present the simulation results for model inference. Figure \ref{sim:fig1} illustrates the estimated functional coefficients obtained using three different methods, along with their corresponding 95\% point-wise confidence intervals. The results indicate that pooling data from both target and source cohorts leads to biased estimates. However, due to the larger combined sample size, this method achieves the narrowest confidence intervals. In contrast, the method using only target cohort data produces unbiased estimates but suffers from wider confidence intervals because of the limited target sample size. Our \texttt{SITL} method, on the other hand, demonstrates a balance between these two extremes. While showing trends similar to those from the target-only method, it effectively incorporates information from source cohorts, resulting in narrower confidence intervals.

\section{Application on Kidney Transplant Data}\label{application}
The motivation for our application case stems from a nationwide cohort study conducted across transplant centers in the United States, focusing on kidney transplant data. The objective of this study is to examine the relationship between long-term survival outcomes of kidney transplant patients and their post-transplant renal function. The human body typically has two kidneys, with one kidney being sufficient to meet normal metabolic demands. However, when both kidneys fail, kidney transplantation becomes the most effective treatment option. Despite its benefits, not all patients can tolerate the surgery or the high doses of steroids and immunosuppressive medications required afterward. Assessing a patient’s suitability for kidney transplantation and predicting post-surgical outcomes are therefore critical steps in the treatment process.

Analyzing long-term survival outcomes using historical data is essential for making informed decisions. However, patient data from different transplant centers are often non-interoperable, making seamless data exchange difficult. As a result, the information available at each center is limited, with smaller centers sometimes lacking historical cases altogether, particularly for patients with complex or rare conditions. To address these challenges, it is vital to incorporate data from other transplant centers to enhance the analysis. By leveraging information across centers, we can improve the robustness and accuracy of insights, ultimately supporting better decision-making in kidney transplantation cases.
  
In this application, we use the dataset that comes from the United Network for Organ Sharing/Organ Procurement Transplantation Network (UNOS/OPTN) as of September 2020. To evaluate the kidney function, an accurate assessment is achieved through the measurement of the glomerular filtration rate (GFR), which integrates creatinine levels along with relevant factors such as age, race, gender, and body size. We set the post-transplant GFR in the first five years as the functional predictor $X(\cdot)$, and define 
the failure time $T$ as the survival time (in years) of kidney recipients starting from five years after the transplant surgery. We excluded patients who underwent multiple kidney transplants or were lost to follow-up within five years. 

We select transplant centers with center codes $05890$, $09021$, and $14415$ as target cohorts, with sample sizes of $226$, $206$, and $203$, respectively. We use 70\% of the data as the training set and 30\% as the test set to evaluate the estimation and prediction performance of our method. For each target cohort, we select four source centers, each with a sample size of at least $600$, to provide additional information. 
\begin{table}[htbp]
\centering
  \fontsize{10}{12}\selectfont
  \begin{threeparttable}
  \caption{The prediction errors for three target centers with varying sample sizes $n_0$ and censoring rates $C_0$  
   using our proposed Similarity-Informed Transfer Learning (\texttt{SITL}) method and three alternative conventional methods including (i) using the target cohort data only (called \texttt{Target}), (ii) pooling target and source cohorts together (called \texttt{Pooled}), and (iii) the hard-threshold transfer learning method (called \texttt{Transfer\_HT}).}
  \label{app:tab1}
  \renewcommand\arraystretch{0.8}
\begin{tabular}{ccccc}
\toprule
\multirow{2}{*}{}&\multirow{3}{*}{\textbf{Methods}}&\multicolumn{3}{c}{\textbf{Target Transplant Center}}\\
\cline{3-5}
 
  &  & \textbf{Center: $05890$} & \textbf{Center: $09021$} & \textbf{Center: $14415$} \\
  &  & \textbf{ ($n_0=226$, $C_0=48.8\%$)} & \textbf{ ($n_0=203$, $C_0=30.0\%$)} & \textbf{($n_0=206$, $C_0=27.3\%$)} \\
  \midrule
\multirow{4}{*}{}& 
 \textbf{Target} 
 & 0.410
 & 2.036
 & 0.450
   \\ 
 & \textbf{Pooled} 
 & 0.204 
 & 0.368
 & 0.376
    \\ 
  & \textbf{Transfer\_HT} 
  & 0.204
  & 0.397
  & 0.346
     \\ 
  & \textbf{\texttt{SITL}} 
  & 0.102
  & 0.312
  & 0.334 \\
  
   \bottomrule
\end{tabular}
\end{threeparttable}
\end{table}

Table \ref{app:tab1} summarizes the prediction errors for the three target center cohorts using our \texttt{SITL} method and three alternative estimation methods. Across all three datasets, the \texttt{SITL} method consistently achieves the lowest prediction errors, with improvements of up to 85\% compared to the method that relies solely on target cohort data. For Center 05890, which has the highest censoring rate (48.8\%), the \texttt{SITL} method reduces prediction errors by 50\% compared to the method that pools target and source cohort data. For Center 14415, with the lowest censoring rate (27.3\%), the reduction is 11\%. When compared to the hard-threshold transfer learning method, the \texttt{SITL} method yields 50\% and 21\% smaller prediction errors for Centers 05890 and 09021, respectively, both of which have high censoring rates. For Center 14415, which has a relatively low censoring rate, the \texttt{SITL} method achieves a 3.5\% reduction in prediction errors.
\begin{figure}[htbp]
	\centering
	\subfigure[Target, $\tau=0.3$]{\includegraphics[width=4.5cm]{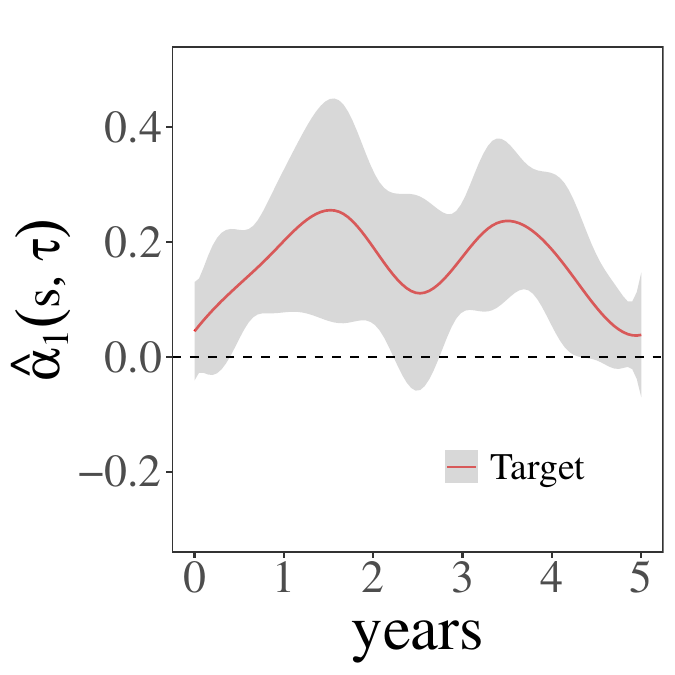}}
	\subfigure[Pooled, $\tau=0.3$]{\includegraphics[width=4.5cm]{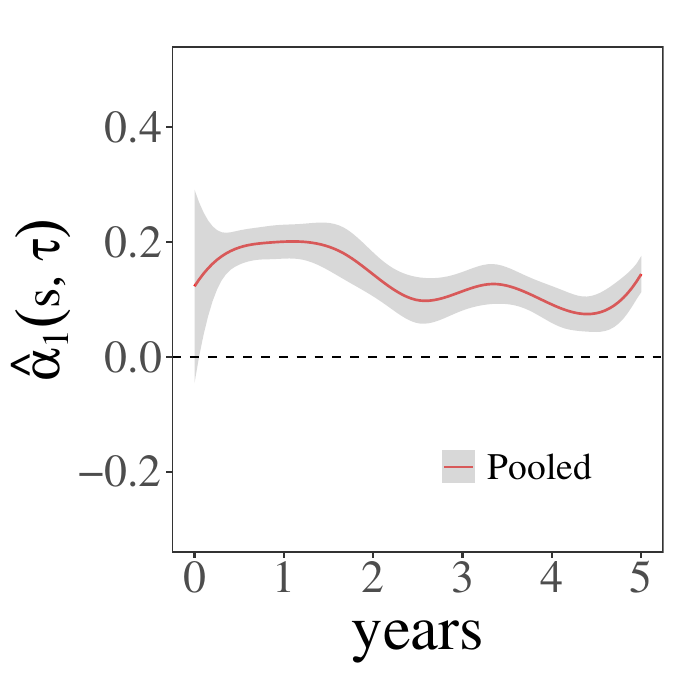}}
    \subfigure[\texttt{SITL}, $\tau=0.3$]{\includegraphics[width=4.5cm]{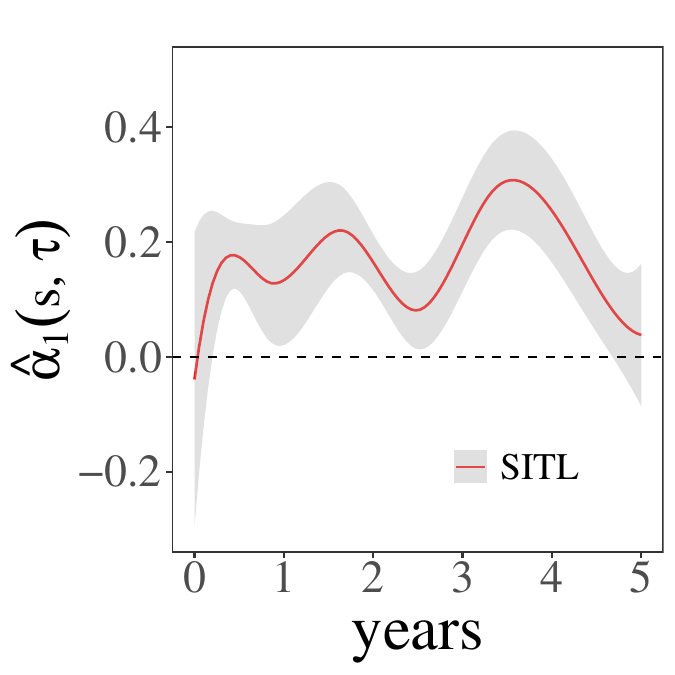}}
	\\
    \subfigure[Target, $\tau=0.5$]{\includegraphics[width=4.5cm]{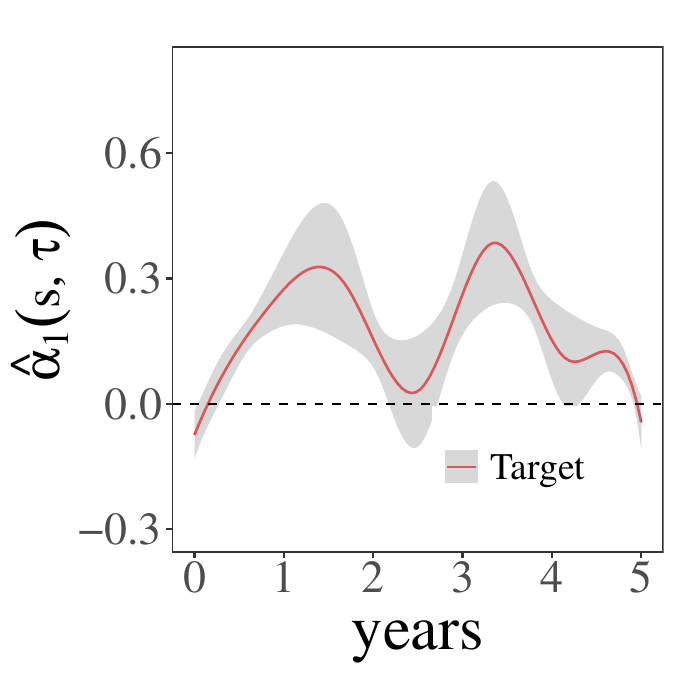}}
    \subfigure[Pooled, $\tau=0.5$]{\includegraphics[width=4.5cm]{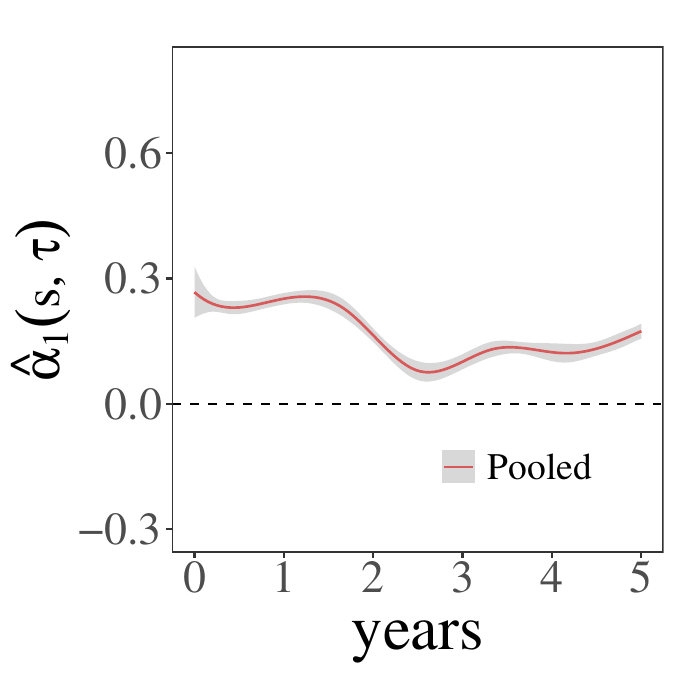}}
	\subfigure[\texttt{SITL}, $\tau=0.5$]{\includegraphics[width=4.5cm]{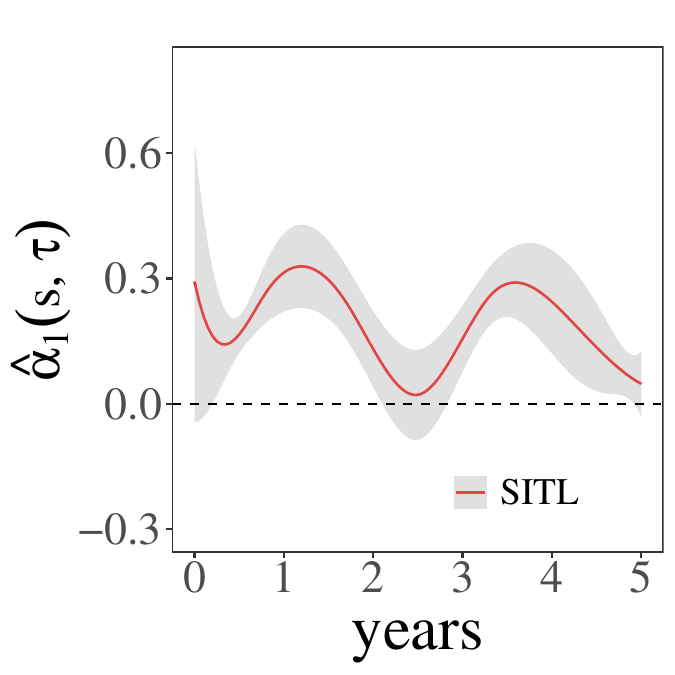}}
    \\
    \subfigure[Target, $\tau=0.7$]{\includegraphics[width=4.5cm]{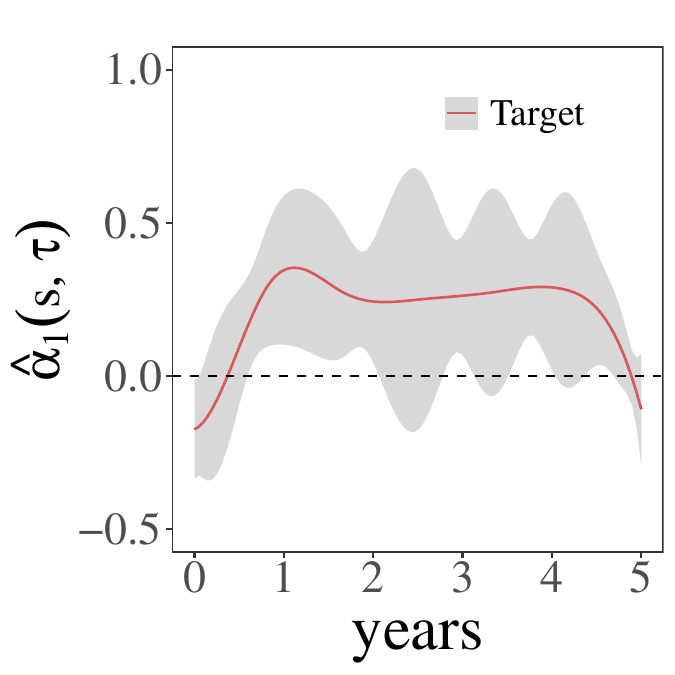}}
	\subfigure[Pooled, $\tau=0.7$]{\includegraphics[width=4.5cm]{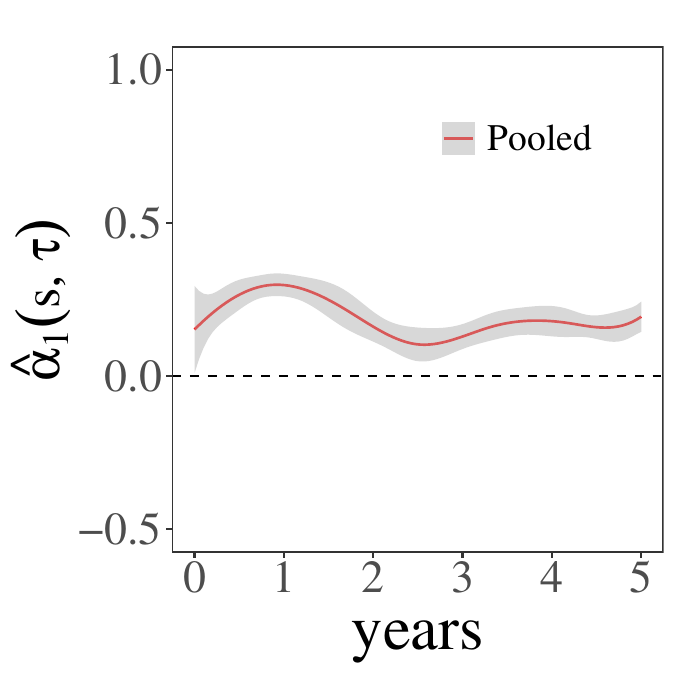}}
    \subfigure[\texttt{SITL}, $\tau=0.7$]{\includegraphics[width=4.5cm]{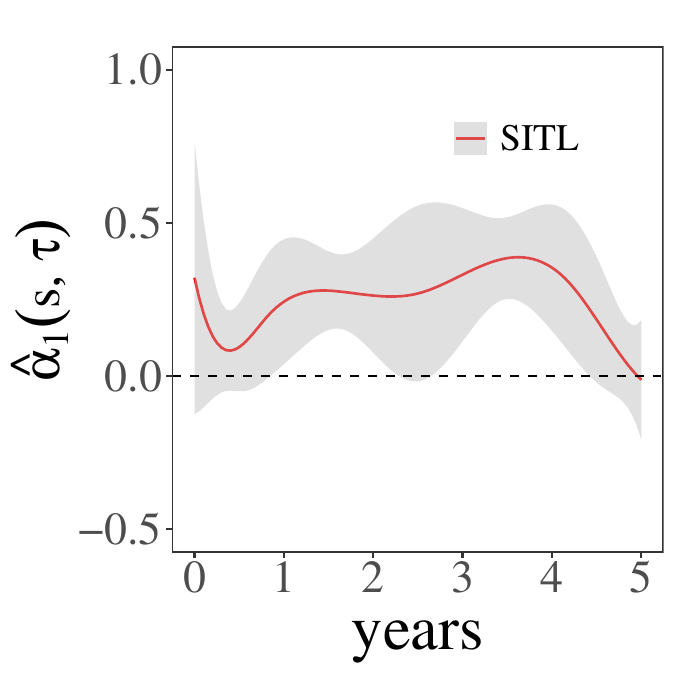}}
	\caption{The estimated functional coefficients  and the corresponding 95\% point-wise confidence intervals for the target Transplant Center $05890$ when the quantile $\tau=0.3,0.5$ and $0.7$, using our proposed Similarity-Informed Transfer Learning (\texttt{SITL}) method and two alternative conventional methods including (i) using the target cohort data only (called \texttt{Target}), and (ii) pooling target and source cohorts together (called \texttt{Pooled}).}\label{app:fig1}
\end{figure}

Figure \ref{app:fig1} shows the estimated functional coefficients and corresponding 95\% point-wise confidence intervals for the transplant center $05890$ dataset under the quantile $\tau=0.3,0.5,0.7$, using the three available methods. The results for the remaining two transplant centers can be found in the supplementary document. First, we observe that, due to the biased estimates produced by the method pooling the source and target cohort data directly, the estimated coefficient curves it obtains differ significantly from those obtained by the SITL method and the method using only the target cohort data, while the curves produced by the latter two methods are quite similar. Our SITL method leverages information provided by the source cohorts and obtains narrower confidence intervals compared to those produced by the method uses only the target dataset. At the 0.3 and 0.5 quantile levels, the estimated functional coefficient curves exhibit a ``hump-shaped" pattern, with peaks around 1 and 4 years post-transplant and a trough near 2.5 years. This suggests that, for patients with shorter survival times beyond five years post-transplant, GFR levels immediately following transplantation and those nearing the five-year mark have a greater influence on survival. This observation aligns with medical theory, which highlights the critical role of immediate post-operative recovery and emphasizes the importance of monitoring physical indicators leading up to the five-year mark in clinical assessments. In contrast, for patients with longer survival times beyond five years (at the 0.7 quantile level), the impact of GFR is more evenly distributed between 1 and 4 years post-transplant.

\section{Conclusions and Discussion}
\label{conclu}
We propose the similarity-informed transfer learning method within the context of multivariate functional censored quantile regression. This approach effectively leverages information from source cohorts to enhance estimation and inference accuracy for target cohorts with limited data, while adhering to data-sharing privacy constraints. Theoretical analyses confirm the statistical efficiency of the method.
In future research, 
we can also develop transfer learning methods for functional censored quantile regression with high-dimensional functional covariates with the variable selection capacity.

\section*{Supplementary Materials}
The supplementary document includes the procedure for the hard-threshold transfer learning method,
additional results from simulation studies, and the theoretical proofs for the theorems in the manuscript. Additionally, we also provide the computing codes for the simulation studies.

\section*{Acknowledgments}
Dr Liu's research was supported by 
National Natural Science Foundation of China (NSFC) (No.12201487) and the Project funded by China Postdoctoral Science Foundation (No.2022M722544).
Dr You's research was supported by 
the National Natural Science Foundation of China (NSFC) (No.11971291) and Innovative Research Team of Shanghai University of Finance and Economics. Dr Cao's research was supported by the Natural Sciences and Engineering Research Council of Canada (NSERC) Discovery
grant (RGPIN-2023-04057) and the Canada Research Chair program. The kidney transplant data set was supported in part by Health Resources and Services Administration contract 234-2005-370011C. 
The content about this data set is the responsibility of the authors alone and does not necessarily reflect the views or policies of the Department of Health and Human Services, nor does mention of trade names, commercial products, or organizations imply endorsement by the U.S. Government.

\bibliographystyle{apalike}
\bibliography{ref,FDA}
\end{document}